\documentclass[9pt,twocolumn,twoside]{opticajnl}
\journal{opticajournal} 

\setboolean{shortarticle}{false}

\usepackage{lineno}

\usepackage{amssymb}
\usepackage{gensymb}

\usepackage{textgreek} 
\usepackage{textcomp}
\usepackage{trsym} 
\usepackage[all]{nowidow}

\usepackage{longtable}  
\usepackage{acro}[=v2]

\usepackage{pdfpages}

\usepackage{tikz}
\usetikzlibrary{plotmarks}
\usetikzlibrary{calc}
\usepackage{pgfplots}
\pgfplotsset{compat=newest}
\pgfplotsset{plot coordinates/math parser=false}
\newlength\figureheight
\newlength\figurewidth 
\usetikzlibrary{positioning}
\acsetup{mark-as-used= first} 

\DeclareAcronym{OAWM}{
    short = OAWM,
    long = optical arbitrary waveform measurement,
    }

\DeclareAcronym{WDM}{
    short = WDM,
    long = wavelength-division multiplexed,
    }

\DeclareAcronym{IQ}{
    short = IQ,
    long = in-phase and quadrature,
    short-indefinite = an,
    long-indefinite = an,
    }

\DeclareAcronym{IQR}{
    short = IQR,
    long = in-phase and quadrature receiver,
    short-indefinite = an,
    long-indefinite = an,
    }
    
\DeclareAcronym{ADC}{
    short = ADC,
    long = analog-to-digital converter,
    short-indefinite = an,
    long-indefinite = an,
    }

\DeclareAcronym{PE-ADC}{
    short = PE-ADC,
    long = photonic-electronic analog-to-digital converter,
    }

 \DeclareAcronym{PE-DAC}{
    short = PE-DAC,
    long = photonic-electronic digital-to-analog converter,
    }

\DeclareAcronym{DSP}{
    short = DSP,
    long = digital signal processing,
    }

\DeclareAcronym{SiP}{
    short = SiP,
    long = silicon photonics,
    }

\DeclareAcronym{InP}{
    short = InP,
    long = indium phosphide,
    short-indefinite = an,
    long-indefinite = an,
    }

\DeclareAcronym{SiN}{
    short =  SiN, 
    long = silicon nitride,
    }

\DeclareAcronym{PIC}{
    short =  PIC, 
    long = photonic integrated circuit,
    }

\DeclareAcronym{IC}{
    short =  IC, 
    long = integrated circuit,
    }

\DeclareAcronym{PWB}{
    short =  PWB, 
    long = photonic wire bond,
    }
  
\DeclareAcronym{AWG}{
    short = AWG,
    long = arbitrary waveform generator,
    short-indefinite = an,
    long-indefinite = an,
    }

\DeclareAcronym{DAC}{
    short = DAC,
    long = digital-to-analog converter,
    }

\DeclareAcronym{CROW}{
    short = CROW,
    long = coupled-resonator optical waveguide,
    }

\DeclareAcronym{RF}{
    short = RF,
    long = radio frequency,
    short-indefinite = an,
    long-indefinite = a,
    }

\DeclareAcronym{CW}{
    short = CW,
    long = continuous wave,
    }

\DeclareAcronym{LO}{
    short = LO,
    long = local oscillator,
    short-indefinite = an,
    long-indefinite = a,
    }

\DeclareAcronym{OH}{
    short = OH,
    long = optical hybrid,
    short-indefinite = an,
    long-indefinite = an,
    }

\DeclareAcronym{VOA}{
    short = VOA,
    long = variable optical attenuator,
    }

\DeclareAcronym{VODL}{
    short = VODL,
    long = variable optical delay line,
    }

\DeclareAcronym{BPD}{
    short = BPD,
    long = balanced photodetector,
    }

\DeclareAcronym{FSR}{
    short = FSR,
    long = free spectral range,
    short-indefinite = an,
    long-indefinite = a,
    }

\DeclareAcronym{OCNR}{
    short = OCNR,
    long = optical carrier-to-noise ratio,
    short-indefinite = an,
    long-indefinite = an,
    }

\DeclareAcronym{CNR}{
    short = CNR,
    long = carrier-to-noise ratio,
    }

\DeclareAcronym{OSNR}{
    short = OSNR,
    long = optical signal-to-noise ratio,
    short-indefinite = an,
    long-indefinite = an,
    }

\DeclareAcronym{SNR}{
    short = SNR,
    long = signal-to-noise ratio,
    }

\DeclareAcronym{CSNR}{
    short = CSNR,
    long = constellation signal-to-noise ratio,
    }

\DeclareAcronym{SINAD}{
    short = SINAD,
    long = signal-to-noise-and-distortion ratio,
    }

\DeclareAcronym{ENOB}{
    short = ENOB,
    long = effective number of bits,
    short-indefinite = an,
    long-indefinite = an,
    long-plural =,
    short-plural =,
    }

\DeclareAcronym{SNDR}{
    short = SNDR,
    long = signal-to-noise-and-distortion ratio,
    }

\DeclareAcronym{EVM}{
    short = EVM,
    long = error vector magnitude,
    short-indefinite = an,
    long-indefinite = an,
    }

\DeclareAcronym{PAPR}{
    short = PAPR,
    long = peak-to-average power ratio ,
    }

\DeclareAcronym{ASE}{
    short = ASE,
    long = amplified spontaneous emission,
    short-indefinite = an,
    long-indefinite = an,
    }
    
\DeclareAcronym{OR}{
    short = OR,
    long = overlap region,
    }

\DeclareAcronym{ORW}{
    short = ORW,
    long = optical reference waveform,
    short-indefinite = an,
    long-indefinite = an,
    }

\DeclareAcronym{FCG}{
    short = FCG,
    long = frequency comb generator,
    }

\DeclareAcronym{OFC}{
    short = OFC,
    long = optical frequency comb,
    short-indefinite = an,
    long-indefinite = an,
    }

\DeclareAcronym{EDFA}{
    short = EDFA,
    long = erbium-doped fiber amplifier,
    short-indefinite = an,
    long-indefinite = an,
    }

\DeclareAcronym{ECL}{
    short = ECL,
    long = external-cavity laser,
    short-indefinite = an,
    long-indefinite = an,
    }

\DeclareAcronym{DKS}{
    short = DKS,
    long = dissipative Kerr soliton,
    }

\DeclareAcronym{BP}{
    short = BP,
    long = band-pass filter,
    }

\DeclareAcronym{RBW}{
    short = RBW,
    long = resolution bandwidth,
    }
    
\DeclareAcronym{WSS}{
    short = WSS,
    long = wavelength-selective switch,
    long-plural = es,
    short-plural = }
    
\DeclareAcronym{PCB}{
    short = PCB,
    long = printed circuit board,
    }

\DeclareAcronym{TIS}{
    short = TIS,
    long = time interleaved sampling,
    }
\DeclareAcronym{DBI}{
    short = DBI,
    long = digital bandwidth interleaving,
    }
\DeclareAcronym{ATI}{
    short = ATI,
    long = asynchronous time interleaving,
    short-indefinite = an,
    long-indefinite = an,
    }

\DeclareAcronym{MZM}{
    short = MZM,
    long = Mach-Zehnder modulator,
    short-indefinite = an,
    long-indefinite = a,
    }

\DeclareAcronym{IQM}{
    short = IQM,
    long = in-phase and quadrature modulator,
    short-indefinite = an,
    long-indefinite = an,
    }

\DeclareAcronym{PM}{
    short = PM,
    long = phase modulator,
    }

\DeclareAcronym{FROG}{
    short = FROG,
    long = frequency-resolved optical gating,
    }

\DeclareAcronym{DFT}{
    short = DFT,
    long = discrete Fourier transform,
    }

\DeclareAcronym{PDF}{
    short = PDF,
    long = probability density functions,
    }

\DeclareAcronym{RMS}{
    short = RMS,
    long = root-mean-square,
    }

\DeclareAcronym{GSG}{
    short = GSG,
    long = ground-signal-ground,
    }

\DeclareAcronym{BER}{
    short = BER,
    long =  bit-error ratio,
    }

\DeclareAcronym{FEC}{
    short = FEC,
    long =  forward-error correction,
    }

\DeclareAcronym{OSA}{
    short = OSA,
    long =  optical spectrum analyzer,
    short-indefinite = an,
    long-indefinite = an,
    }

\DeclareAcronym{TFLN}{
    short = TFLN,
    long =  thin-film lithium-niobate,
    }

\DeclareAcronym{EO}{
    short = EO,
    long =  electro-optic,
    short-indefinite = an,
    long-indefinite = an,
    }
    
\DeclareAcronym{OE}{
    short = OE,
    long =  opto-electronic,
    short-indefinite = an,
    long-indefinite = an,
    }
    
\DeclareAcronym{EOE}{
    short = EOE,
    long =  electro-optic-electric,
    short-indefinite = an,
    long-indefinite = an,
    }

\DeclareAcronym{OAWG}{
    short = OAWG,
    long =  optical arbitrary waveform generation,
    short-indefinite = an,
    long-indefinite = an,
    }

\DeclareAcronym{QAM}{
    short = QAM,
    long =  quadrature amplitude modulation,
    }

\DeclareAcronym{PAM}{
    short = PAM,
    long =  pulse-amplitude modulation,
    }

\DeclareAcronym{MMI}{
    short = MMI,
    long = multi-mode interference,
    short-indefinite = an,
    long-indefinite = a,
    }
    
\DeclareAcronym{ER}{
    short = ER,
    long = extinction ratio,
    }
    
   \DeclareAcronym{IL}{
    short = IL,
    long = insertion loss,
    }

\DeclareAcronym{FOM}{
    short = FOM,
    long = figure of merit,
    }

\DeclareAcronym{POH}{
    short = POH,
    long = plasmonic organic hybrid,
    }

\DeclareAcronym{SOH}{
    short = SOH,
    long = silicon organic hybrid,
    }

\DeclareAcronym{CC-SOH}{
    short = CC-SOH,
    long = capacitively coupled silicon organic hybrid,
    }

\DeclareAcronym{SSBI}{
    short = SSBI,
    long = signal-signal beat interference,
    }

 \DeclareAcronym{LOSPR}{
    short = LOSPR,
    long = local-oscillator-to-signal power ratio,
    } 
    
 \DeclareAcronym{FSIR}{
    short = FSIR,
    long = full-scale input range,
    }

 \DeclareAcronym{CMRR}{
    short = CMRR,
    long = common-mode rejection ratio,
    }   

 \DeclareAcronym{FMCW}{
    short = FMCW,
    long = frequency-modulated continues-wave,
    }

\DeclareAcronym{CEO}{
   short = CEO,
   long = carrier-envelope offset,
   }    

\DeclareAcronym{PSD}{
   short = PSD,
   long = power spectral density,
   }

\DeclareAcronym{FWHM}{
   short = FWHM,
   long = full width at half maximum,
   }  

\DeclareAcronym{RIN}{
   short = RIN,
   long = relative intensity noise,
   }  

\DeclareAcronym{FWM}{
   short = FWM,
   long = four-wave mixing,
   }

\DeclareAcronym{EB}{
   short = EB,
   long = exabytes,
   }  

\DeclareAcronym{CAGR}{
   short = CAGR,
   long = compound annual growth rate}  

\DeclareAcronym{TDM}{
   short = TDM,
   long = time division multiplexed,
   }  

\DeclareAcronym{CMOS}{
   short = CMOS,
   long = complementary metal-oxide semiconductor,
   }  

\DeclareAcronym{BiCMOS}{
   short = BiCMOS,
   long = bipolar \acs{CMOS},
   }  

\DeclareAcronym{AMUX}{
   short = AMUX,
   long = analog multiplexer,
   }  

\DeclareAcronym{FI}{
   short = FI,
   long = frequency-interleaving,
   }

\DeclareAcronym{DEMUX}{
   short = DEMUX,
   long = demultiplexing filter, 
    long-plural = s,
   short-plural = es,
   }

\DeclareAcronym{MUX}{
   short = MUX,
   long = multiplexer,
   long-plural = s,
   short-plural = es,
   }  

\DeclareAcronym{OTDM}{
short = OTDM,
long = optical time-division multiplexing,
long-plural = s,
short-plural = es,
}

\DeclareAcronym{SMF}{
short = SMF,
long = single mode fiber,
}

\DeclareAcronym{NSR}{
short = NSR,
long = noise-to-signal ratio,
}

\DeclareAcronym{NRE}{
short = NRE,
long = non-recurring engineering,
}

\DeclareAcronym{TandM}{
short = T\&M,
long = test and measurement,
}  

\DeclareAcronym{Ge}{
short = Ge,
long = germanium,
}

\DeclareAcronym{SCT}{
short = SCT,
long = signal-combining tree,
}  

\DeclareAcronym{SCE}{
short = SCE,
long = signal-combining element,
}  

\DeclareAcronym{CLK}{
short = CLK,
long = clock,
}  

\DeclareAcronym{PS}{
short = PS,
long = phase shifter,
}  

\DeclareAcronym{PI}{
short = PI,
long = proportional-integral,
}

\DeclareAcronym{RRC}{
short = RRC,
long = root-raised cosine,
}

\DeclareAcronym{TIA}{
short = TIA,
long = transimpedance amplifier,
}

\DeclareAcronym{LP}{
short = LP,
long = low-pass filter,
short-indefinite = an,
long-indefinite = a,
}

\DeclareAcronym{SOA}{
short = SOA,
long = semiconductor optical amplifier,
short-indefinite = an,
long-indefinite = a,
} 

\newcommand{\ASnut}[1][\nu]{{\underline{{A}}_{\mathrm{S},#1}}(t)}
\newcommand{\ASnuf}[1][\nu]{{\underline{\tilde{A}}_{\mathrm{S},#1}}(f)}

\newcommand{\aSt}{{\underline{{a}}_{\mathrm{S}}(t)}}

\newcommand{\aSf}{{\underline{\tilde{a}}_{\mathrm{S}}(f)}}

\newcommand{\aSnut}[1][\nu]{{\underline{{a}}_{\mathrm{S},#1}}(t)}
\newcommand{\aSnuf}[1][\nu]{{\underline{\tilde{a}}_{\mathrm{S},#1}}(f)}

\newcommand{\Inuf}[1][\nu]{{\tilde{I}_{#1}(f)}}
\newcommand{\Qnuf}[1][\nu]{{\tilde{Q}_{#1}(f)}}
\newcommand{\InufConj}[1][\nu]{{\tilde{I}^*_{#1}(-f)}}
\newcommand{\QnufConj}[1][\nu]{{\tilde{Q}^*_{#1}(-f)}}
\newcommand{\Inut}[1][\nu]{{I_{#1}(t)}}
\newcommand{\Qnut}[1][\nu]{{Q_{#1}(t)}}

\newcommand{\HnuIf}[1][\nu]{{\underline{\tilde{H}}_{#1}^{\mathrm{(I)}}\!(f)}}
\newcommand{\HnuIfConj}[1][\nu]{{\underline{\tilde{H}}_{#1}^{\mathrm{(I)}*}\!(-f)}}
\newcommand{\HnuQf}[1][\nu]{{\underline{\tilde{H}}_{#1}^{\mathrm{(Q)}}\!(f)}}
\newcommand{\HnuQfConj}[1][\nu]{{\underline{\tilde{H}}_{#1}^{\mathrm{(Q)}*}\!(-f)}}

\newcommand{\nuRange}[1][N]{{\nu=1,\ldots,#1}}

\newcommand{\symACircRed}[1][1]{\raisebox{-0.8mm}{\includegraphics[scale=#1]{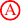}}}

\newcommand{\symBCircRed}[1][1]{\raisebox{-0.8mm}{\includegraphics[scale=#1]{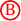}}}

\newcommand{\symCCircRed}[1][1]{\raisebox{-0.8mm}{\includegraphics[scale=#1]{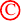}}}

\newcommand{\symDCircRed}[1][1]{\raisebox{-0.8mm}{\includegraphics[scale=#1]{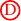}}}

\newcommand{\symDCircBlue}[1][1]{\raisebox{-0.8mm}{\includegraphics[scale=#1]{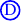}}}
\newcommand{\symECircRed}[1][1]{\raisebox{-0.8mm}{\includegraphics[scale=#1]{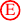}}}
\newcommand{\symFCircRed}[1][1]{\raisebox{-0.8mm}{\includegraphics[scale=#1]{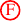}}}

\definecolor{color_slice_1}{rgb}{1, 0, 0}
\definecolor{color_slice_2}{rgb}{1,0.75,0}
\definecolor{color_slice_3}{rgb}{0, 0.8, 0}
\definecolor{color_slice_4}{rgb}{0.49,0.18,0.56}

\definecolor{color_slice_1_txt}{rgb}{1, 0, 0}
\definecolor{color_slice_2_txt}{rgb}{0.84,0.44,0}
\definecolor{color_slice_3_txt}{rgb}{0, 0.5, 0}
\definecolor{color_slice_4_txt}{rgb}{0.49,0.18,0.56}

\definecolor{color_Supp1}{rgb}{0,0,0}

\newcommand{\subfigcap}[1]{\unskip\quad\textbf{#1}}

\title{Optical Arbitrary Waveform Generation (OAWG) Using Actively Phase-Stabilized Spectral Stitching}

\author[1,2,+,*]{Daniel~Drayss}
\author[1,+]{Dengyang~Fang}
\author[1,+]{Alban~Sherifaj}
\author[1]{Huanfa~Peng}
\author[1]{Christoph~F\"ullner}
\author[3]{Thomas~Henauer}
\author[4]{Grigory~Lihachev}
\author[1]{Tobias~Harter}
\author[1]{Wolfgang~Freude}
\author[1]{Sebastian~Randel}
\author[4]{Tobias~J.~Kippenberg}
\author[3]{Thomas~Zwick}
\author[1,2,**]{Christian~Koos}

\affil[1]{Institute of Photonics and Quantum Electronics (IPQ), Karlsruhe Institute of Technology (KIT), 76131 Karlsruhe, Germany}
\affil[2]{Institute of Microstructure Technology (IMT), Karlsruhe Institute of Technology (KIT), 76344 Eggenstein-Leopoldshafen, Germany}
\affil[3]{Institute of Radio Frequency Engineering and Electronics (IHE), Karlsruhe Institute of Technology (KIT), 76131 Karlsruhe, Germany}
\affil[4]{Institute of Physics, Swiss Federal Institute of Technology Lausanne (EPFL), CH-1015 Lausanne, Switzerland}
\affil[+]{Contributed equally. \quad \textsuperscript{*}daniel.drayss@kit.edu. \quad \textsuperscript{**}christian.koos@kit.edu}

\acresetall 
\begin{abstract}
The conventional way of generating optical waveforms relies on the \acf{IQ} modulation of a \acf{CW} laser tone. In this case, the bandwidth of the resulting optical waveform is limited by the underlying electronic components, in particular by the \acfp{DAC} generating the drive signals for the \acs{IQ} modulator. This bandwidth bottleneck can be overcome by using a concept known as \acf{OAWG}, where multiple \acs{IQ} modulators and \acsp{DAC} are operated in parallel to first synthesize individual spectral slices, which are subsequently combined to form a single ultra-broadband arbitrary optical waveform. 
However, targeted synthesis of arbitrary optical waveforms from multiple spectral slices has so far been hampered by difficulties to maintain the correct optical phase relationship between the slices. 
In this paper, we propose and demonstrate spectrally sliced \acs{OAWG} with active phase stabilization, which permits targeted synthesis of truly arbitrary optical waveforms. We demonstrate the viability of the scheme by synthesizing optical waveforms with record-high bandwidths of up to 325\,GHz from four individually generated optical tributaries. 
In a proof-of-concept experiment, we use the \acs{OAWG} system to generate 32QAM data signals at symbol rates of up to 320\,GBd, which we transmit over 87\,km of single-mode fiber and receive by a two-channel non-sliced \acf{OAWM} system, achieving excellent signal quality. 
We believe that our scheme can unlock the full potential of \acs{OAWG} and disrupt a wide range of applications in high-speed optical communications, photonic-electronic digital-to-analog conversion, as well as advanced test and measurement in science and industry.
\end{abstract}

\setboolean{displaycopyright}{false} 

\begin{document}

\maketitle

\acresetall 
\section{Introduction}
\label{sec:OAWG_Introduction}
Optical arbitrary waveforms are usually generated by sending a \ac{CW} optical carrier through an \ac{IQM}\acuse{IQ}, which is driven by a pair of \ac{RF} signals \cite{Kikuchi_2016_JLT}. In this case, the bandwidth of the resulting optical waveform is dictated by the bandwidth of the underlying electronic components such as the \acp{DAC}, the driver amplifiers, and the \acp{IQM}, which are typically limited to less than 100\,GHz \cite{Chen_2017_JLT, Schmidt_2018_OE, Drenski_2018_OFC, Kim_2022_JSSS_224Gbs_DAC, Li_2020_IEEEAccess, ElAassar_2020_IEEEMWCL, SHF_2023_datasheetT850B, Xu_2022_Optica}. 
This limitation can be overcome by using \ac{OAWG} techniques \cite{Fontaine_2010_OE,Geisler_2011_OE, Geisler_2011_PhotonJ, RiosMuller_2015_OFC}, which exploit optical frequency combs as multi-wavelength carriers for spectrally sliced signal synthesis. In this approach, the phase-locked comb tones are first modulated independently by an array of \acp{IQM} and associated \acp{DAC}, and the resulting tributary signals are then merged into a single broadband optical waveform. However, while this scheme renders the bandwidth of the synthesized optical waveform independent of the bandwidth of the underlying electronic components, synthesis of truly arbitrary waveforms was so far hindered by phase drifts among the individually generated optical tributary signals. Previous experiments thus either required a selection of measurements with coincidentally correct phase relations \cite{Fontaine_2010_OE, Geisler_2011_OE, Geisler_2011_PhotonJ} or, in the case of communication signals, non-standard data-aided processing at the receiver to compensate for the phase drifts among the tributaries at the comb-based \ac{OAWG} transmitter \cite{RiosMuller_2015_OFC}. This severely limits the viability and the application potential of spectrally sliced \ac{OAWG}.

In this paper, we demonstrate an \ac{OAWG} scheme that relies on active stabilization of the phases with which the various tributary signals are combined, thereby enabling the generation of truly arbitrary waveforms with good signal quality \cite{Henauer_2022_OFC, Drayss_2023_CLEO}. In a proof-of-concept experiment, we implement a four-slice phase-stabilized \ac{OAWG} system, offering a record-high bandwidth of 325\,GHz. Our \ac{OAWG} system is carefully calibrated based on a dedicated system model, thus permitting high-fidelity waveform generation. We demonstrate the viability of the concept in a high-symbol-rate optical transmission experiment, combining phase-stabilized \ac{OAWG} with non-sliced \ac{OAWM} \cite{Drayss_2023_Optica}. Using the combined \ac{OAWG}/\ac{OAWM} setup, we transmit fully coherent 16QAM and 32QAM signals with symbol rates of up to 320\,GBd. To the best of our knowledge, our work represents the first \ac{OAWG} demonstration using actively phase-stabilized signal synthesis, leading to the highest bandwidth so far achieved in any \ac{OAWG} experiment as well as to the highest symbol rate demonstrated for fully coherent \acuse{QAM}\ac{QAM} data signals, for which the pulse shape is defined digitally. We believe that our scheme can unlock the full potential of \ac{OAWG} and serve a wide range of applications, comprising high-speed optical communications, microwave and millimeter-wave photonics, photonic-electronic digital-to-analog conversion, or advanced test and measurement in science and industry.

\section{Vision and concept of OAWG transmitter}
\label{sec:OAWG_VisionAndConcept}
\begin{figure*} [htb]
\centering
\includegraphics[width = \linewidth, trim={0cm 0cm 0cm 0cm},clip]{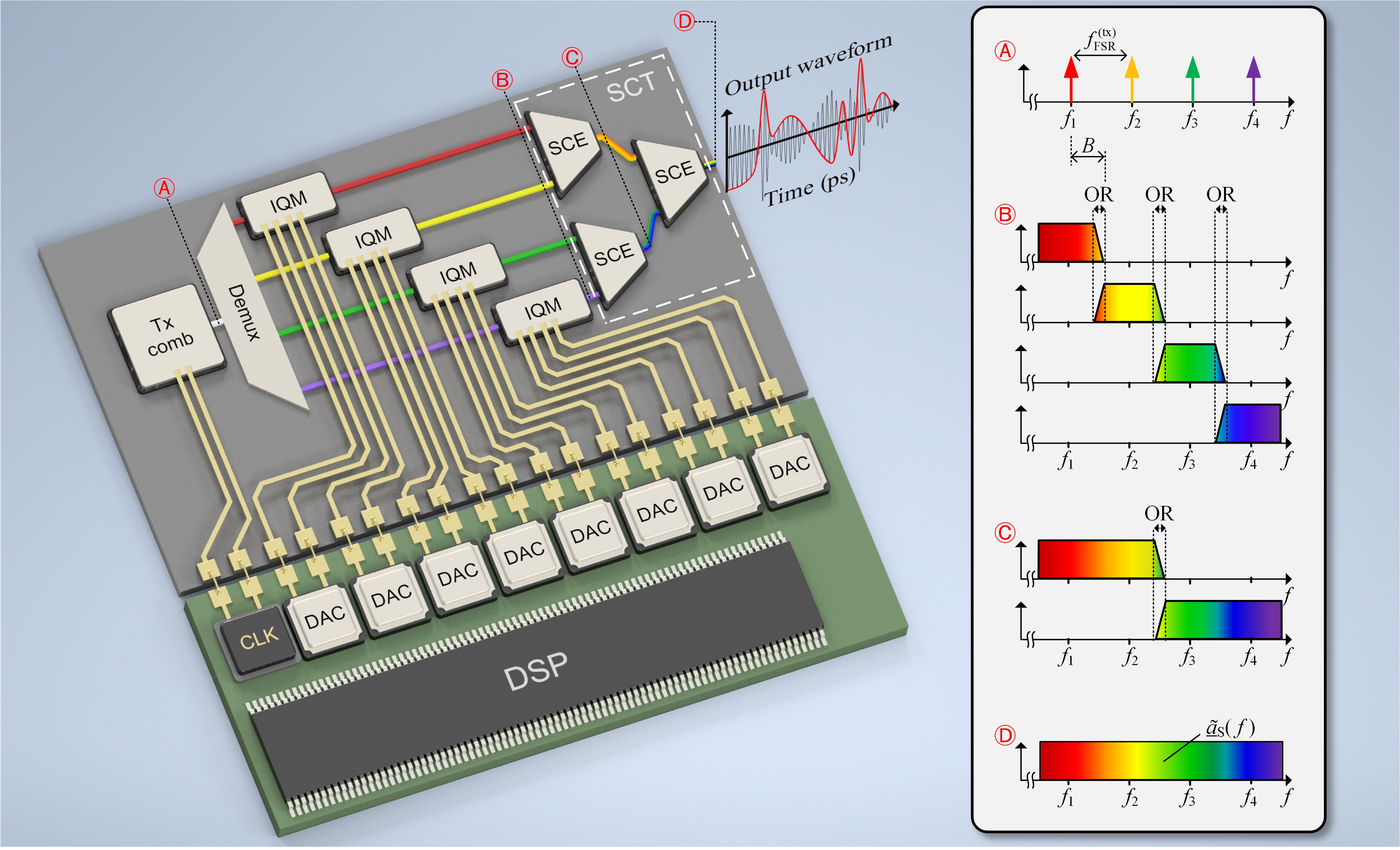}
\captionsetup{font={stretch=1.1, small}}
\caption{\acswitchoff Simplified vision of a spectrally sliced \acs{OAWG} system that comprises $N=4$ channels with active phase stabilization and that allows to generate optical arbitrary waveforms varying on timescales of a few picoseconds. A transmitter frequency comb (Tx comb) is used to generate $N=4$ phase-locked optical tones at frequencies $\color{color_slice_1_txt}f_1,\color{color_slice_2_txt}f_2,\color{color_slice_3_txt}f_3,\color{color_slice_4_txt}f_4$, spaced by a \acf{FSR} $f_{\mathrm{FSR}}^{\mathrm{(tx)}}$, see Inset~\symACircRed[0.9]{}. These tones are separated by a demultiplexing filter (Demux) and subsequently serve as optical carriers for \acf{IQ} modulation. The \acs{IQ} modulators (\acsp{IQM}) are electrically driven by an array of $2N=8$ synchronized \acfp{DAC} having each a bandwidth $B>(f_{\mathrm{FSR}}^{\mathrm{(tx)}}/2)$. The drive signals are calculated based on the desired output waveform using a \acf{DSP} unit. The various output signals of the \acsp{IQM} are combined by a binary \acf{SCT} that consists of $N-1=3$ \acfp{SCE}. Each \acs{SCE} merges two spectrally adjacent tributaries, see Inset~\symCCircRed[0.9]{}, where the slowly varying phase offset $\Delta\varphi(t)$ is minimized by a closed loop control. To this end, the various tributaries are designed to exhibit a slight spectral overlap with their respective neighbor. This leads to interference within well-defined \acfp{OR}, Insets~\symBCircRed[0.9]{} and~\symCCircRed[0.9]{}, which provides the feedback signals for the closed-loop control. The output signal of the last \acs{SCE} corresponds to the desired optical waveform $\aSt$  with spectrum $\aSf$, Inset~\symDCircRed[0.9]{}.}
\label{fig:OAWG_F1_Vision}
\end{figure*}
A simplified vision of an exemplary spectrally sliced \ac{OAWG} system with active phase stabilization is illustrated in Fig.~\ref{fig:OAWG_F1_Vision}. The example relies on $N=4$ tributary signals, which are generated by \acl{IQ} modulation of four phase-locked tones of an optical frequency comb (Tx comb) and which are merged in a binary tree of $N-1=3$ \acp{SCE}, each comprising feed-back-based phase stabilization. The $N=4$ comb tones at frequencies $f_1, f_2, f_3, f_4$, spaced by \iac{FSR} $f_{\mathrm{FSR}}^{\mathrm{(tx)}}$, are illustrated in Inset~\symACircRed{} of Fig.~\ref{fig:OAWG_F1_Vision}. Importantly, the \ac{FSR} of the Tx comb source is synchronized to the \ac{CLK} of the \ac{DAC} array to achieve full coherence among all subsequently generated spectral slices. This can, e.g., be accomplished by generating the Tx comb via electro-optic modulation of a \ac{CW} laser tone where the modulator's driver signals are synchronized to the electronic clock \cite{Parriaux_2020_AOP, Hu_2022_NaturePhotonics} or by \ac{RF} synchronized pulsed solid-state lasers \cite{PilotPhotonics_2023_LyraOcs1000_datasheet, Menhir1550_3p5GHz_2022_datasheet}. The generated frequency comb is then fed to a \ac{DEMUX}, which may be implemented, e.g., as a \ac{WSS}, an arrayed waveguide grating \cite{Fontaine_2011_OptCommun, Fang_2024_OFC}, or a bank of ring filters \cite{Park_2011_OE, Fang_2022_JLT}, and which separates the various comb tones for subsequent \ac{IQ} modulation. The \acp{IQM} are electrically driven by an array of $2N=8$ synchronized \acp{DAC}, which are connected to a \ac{DSP} unit that calculates the various \ac{IQ} drive signals from the targeted output waveform. Merging of the $N$ tributaries in the \ac{SCT} finally produces the targeted broadband arbitrary optical waveform $\aSt$ with spectrum $\aSf$ see Inset~\symDCircRed{} of 
Fig.~\ref{fig:OAWG_F1_Vision}. Within the \ac{SCT}, each of the $N-1=3$ \acp{SCE} combines two adjacent tributaries, see Inset~\symCCircRed{} in Fig.~\ref{fig:OAWG_F1_Vision}, where the slowly varying phase offset $\Delta\varphi(t)$ is minimized by a closed loop control. To this end, the various tributaries are deliberately designed to exhibit a slight spectral overlap with their respective neighbor. This leads to interference within well-defined \acp{OR}, see Insets~\symBCircRed{} and~\symCCircRed{} in Fig.~\ref{fig:OAWG_F1_Vision}, which provides the feedback signals for the closed-loop control, refer to Fig.~\ref{fig:OAWG_F2_SCE_setup} and~\ref{fig:OAWG_F3_SCE_control_on_off} and the discussion thereof in Section~\ref{sec:OAWG_SCE} below for details. The bandwidth of the generated optical arbitrary waveform exceeds that of the individual \acp{DAC} by a factor of $\sim2N$, allowing the synthesis of waveforms with bandwidths of hundreds of GHz that vary on single-digit picosecond time scales.

\subsection{System model}
\label{sec:OAWG_SystemModel}

For high-fidelity \ac{OAWG} using the scheme illustrated in Fig.~\ref{fig:OAWG_F1_Vision}, the characteristics of the various system components must be known and considered in the design of the drive signals that are fed to the various \acp{IQM}. In this section, we derive a linear system model for a generalized \ac{OAWG} transmitter featuring an array of $N$ \acp{IQM} and $N$ corresponding spectrally sliced tributaries. The concept of the active phase stabilization of the \ac{SCE} is explained in the subsequent Section~\ref{sec:OAWG_SCE}. In a first step, we express the spectrum $\aSf$ of the arbitrary optical output waveform  $\aSt$ as a superposition of $N$ individual tributaries $\aSnut$  with spectra $\aSnuf$, $\nuRange$,
\begin{equation}
\aSt = \sum_{\nu=1}^{N}{\aSnut}
\quad \TransformHoriz \quad 
\aSf = \sum_{\nu=1}^{N}{\aSnuf},
\label{eq:OAWG_EQ1_sum_aSnu}
\end{equation}
where the transformation symbol ($\TransformHoriz$) is used to denote a transfer between the time and the frequency domain via a Fourier transform. Note that the tributary signals $\aSnut$ refer to the output of the \ac{SCT}, point \symDCircRed{} in Fig.~\ref{fig:OAWG_F1_Vision}, such that the transfer characteristics of all \acp{SCE} and other system elements are already included. Each of the time-domain tributary signals $\aSnut$ can be expressed by a complex-valued envelope $\ASnut$  and a corresponding comb-tone, acting as a carrier at frequency $f_\nu$,
\begin{equation}
\aSnut = \ASnut \mathrm{e}^{\mathrm{j}2 \pi f_\nu t}
\quad \TransformHoriz \quad 
\aSnuf = \underline{\tilde{A}}_{\mathrm{S},\nu}(f-f_\nu).
\label{eq:OAWG_EQ2_ASnu}
\end{equation}
The complex-valued envelope $\ASnut$ of each of the optical tributary signals is related to the corresponding electrical drive signals $\Inut$ and $\Qnut$  of the various \acp{IQM} by equivalent baseband transfer functions $\HnuIf$ and $\HnuQf$, that combine all optical and electrical transfer functions of the respective signal path,
\begin{equation}
\ASnuf = \HnuIf \Inuf + \mathrm{j} \HnuQf \Qnuf.
\label{eq:OAWG_EQ3_ASnuFRWD}
\end{equation}
In this relation, $\Inuf$ and $\Qnuf$ are the Fourier spectra of the electrical \ac{IQ} drive signals $\Inut$ and $\Qnut$, respectively. Note that for the derivation of the above relation we assumed that the various \acp{IQM} are biased at the zero-transmission point, that the in-phase and quadrature components have an ideal $90\degree$ phase relationship (factor $\mathrm{j}$), and that all drive signals are sufficiently small such that a linear approximation of the \ac{EO} transfer function of the \ac{MZM} can be used.

The \ac{IQ} drive signals $\Inut$ and $\Qnut$ finally need to be designed to produce the targeted arbitrary optical waveform $\underline{a}_{\mathrm{S}}^{\mathrm{(tar)}}(t)$ at the \ac{SCT} output, which is accomplished in two steps: First, we calculate the targeted envelope spectra $\underline{\tilde{A}}_{\mathrm{S},\nu}^{\mathrm{(tar)}}(f)$ of the various tributary signals, $\nuRange$ from the targeted optical waveform $\underline{a}_{\mathrm{S}}^{\mathrm{(tar)}}(t)$. 
In a second step, we use appropriate signal pre-distortion to conceive \ac{IQ} drive signals that compensate for the transfer characteristics of the various electrical and optical components along the signal path to the \ac{SCT} output. For the first step, we multiply the targeted optical spectrum $\underline{\tilde{a}}_{\mathrm{S}}^{\mathrm{(tar)}}(f)$ by a series of slice-specific real-valued optical window functions $w_\nu(f)$, $\nuRange$, to obtain the targeted spectral envelopes $\underline{\tilde{A}}_{\mathrm{S},\nu}^{\mathrm{(tar)}}(f)$ that need to be modulated onto the corresponding comb tones, 
\begin{equation}
\underline{\tilde{A}}_{\mathrm{S},\nu}^{\mathrm{(tar)}}(f-f_\nu)
=\underline{\tilde{a}}_{\mathrm{S},\nu}^{\mathrm{(tar)}}(f)
= w_\nu(f) \times \underline{\tilde{a}}_{\mathrm{S}}^{\mathrm{(tar)}}(f)
\label{eq:OAWG_EQ4_ASnuFtar}
\end{equation}
Note that the optical bandwidth of each window function $w_\nu(f)$ in Eq.~\ref{eq:OAWG_EQ4_ASnuFtar} is chosen slightly bigger than the \ac{FSR} $f_{\mathrm{FSR}}^{\mathrm{(tx)}}$ such that the spectra $\underline{\tilde{a}}_{\mathrm{S},\nu}^{\mathrm{(tar)}}(f)$ of adjacent tributary signals exhibit the desired spectral overlap, see \acs{OR} in Inset~\symBCircRed{} of Fig.~\ref{fig:OAWG_F1_Vision}, and add up to the desired amplitude within the corresponding \acl{OR}, i.e., $\sum_{\nu=1}^{N}w_{\nu}(f)=1$ for all frequencies within the bandwidth of the targeted signal. In the experimental demonstration discussed in Section~\ref{sec:OAWG_Exp} below, we choose an \acl{OR} with a bandwidth of 5\,GHz and a linear decaying window function $w_\nu(f)$, $\nuRange$ see Inset~\symBCircRed{} of Fig.~\ref{fig:OAWG_F1_Vision}. 
Based on the targeted spectral envelopes $\underline{\tilde{A}}_{\mathrm{S},\nu}^{\mathrm{(tar)}}(f)$, we then derive the drive signals of the various \acp{IQM} by appropriate pre-distortion based on the respective transfer functions $\HnuIf$ and $\HnuQf$, thus compensating for the frequency response of the \ac{DAC}-array, the \acp{IQM}, the electrical and the optical amplifiers, and the optical propagation path through the \ac{SCT}. To this end, we first measure all baseband transfer functions $\HnuIf$, $\HnuQf$, in a dedicated calibration measurement, see \textcolor{color_Supp1}{Supplement\,1} Section\,2 for a detailed description of the calibration procedure. We then make use of the symmetry relation for spectra of real-valued signals, $\Inuf = \InufConj$ and $\Qnuf = \QnufConj$, and expand Eq.~\ref{eq:OAWG_EQ3_ASnuFRWD},
\begin{equation}
\begin{bmatrix}
    \underline{\tilde{A}}_{\mathrm{S},\nu}^{\mathrm{(tar)}}(f) \\
    \underline{\tilde{A}}_{\mathrm{S},\nu}^{\mathrm{(tar)}*}(-f)
\end{bmatrix}
=
\begin{bmatrix}
    \HnuIf & \mathrm{j}\HnuQf \\
    \HnuIfConj & -\mathrm{j} \HnuQfConj
\end{bmatrix}
\begin{bmatrix}
    \Inuf \\
    \Qnuf
\end{bmatrix}.
\label{eq:OAWG_EQ5_system_FWRD_matrix}
\end{equation}
The spectra $\Inuf$  and $\Qnuf$ of the pre-distorted \ac{IQ} drive signals $\Inut$ and $\Qnut$ are then obtained by inverting Eq.~\ref{eq:OAWG_EQ5_system_FWRD_matrix}.

\subsection{Actively phase-stabilizing signal-combining element}
\label{sec:OAWG_SCE}
\begin{figure}[htb]
    \centering
    \includegraphics[]{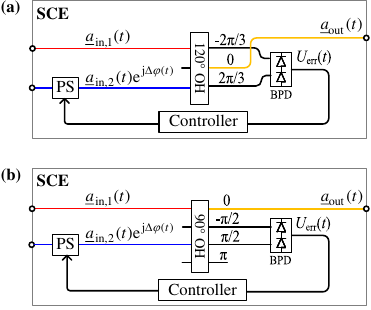}
    \caption{\acswitchoff Implementation of \acfp{SCE} with closed-loop phase stabilization. \subfigcap{(a)}~Conceptual setup of an \acs{SCE} relying on a $120\degree$ \acf{OH}, which is implemented as 3\texttimes3 \acf{MMI} coupler with one unused input port. The \acs{MMI} superimposes the input signals with relative phases of $-2\pi/3$, $0$, and $2\pi/3$ at the various output ports. Two overlapping spectral slices $\underline{a}_{\mathrm{in},1}(t)$ and $\underline{a}_{\mathrm{in},2}(t)\mathrm{exp}\bigl(\mathrm{j}\Delta\varphi(t)\bigr)$ with phase error $\Delta\varphi(t)$ are combined at the “zero-phase” port, leading to the optical output signal $\underline{a}_{\mathrm{out}}(t)$. The error signal $U_{\mathrm{err}}(t)$ is generated by connecting the remaining two output ports of the $120\degree$ \acs{OH}, which are associated with relative phases of $-2\pi/3$ and $2\pi/3$ between the input signals, to a low-speed \acf{BPD}. Upon low-pass filtering, the resulting \acs{BPD} output signal $U_{\mathrm{err}}(t)$ is essentially proportional to phase error $\Delta\varphi(t)$ for a linear approximation close to the desired operating point $\Delta\varphi = 0$, see Eq.~\ref{eq:OAWG_EQ6_Uerr} and Section\,1.1 in \textcolor{color_Supp1}{Supplement\,1}. A \acf{PI} controller is used to drive the \acf{PS} and to compensate for the measured phase error $\Delta\varphi(t)$. \subfigcap{(b)}~Alternative \acs{SCE} implementation using a $90\degree$ \acs{OH} as passive combiner, e.g., implemented as a 4\texttimes4 \acs{MMI} with two unused input ports. The targeted output signal is found again at the “zero-phase” port, and the error signals is derived by a \acs{BPD} connected to the “$-\pi/2$” and the “$\pi/2$” ports. The remaining output port “$\pi$" is unused in this implementation.}
    \label{fig:OAWG_F2_SCE_setup}
\end{figure}

A key aspect for the \ac{OAWG} concept depicted in Fig.~\ref{fig:OAWG_F1_Vision} is the actively phase-stabilized superposition of two slightly overlapping spectrally adjacent slices by the \ac{SCE}. This is not only essential for setups relying on fiber-pigtailed components, which are subject to significant phase variations due to vibrations and thermal drifts, but also for integrated systems, where a closed-loop phase control allows for stable high-fidelity signal generation independent of thermal drifts within the underlying \ac{PIC} \cite{Milvich_2021_AOP} -- just like a bias control on integrated \acp{MZM} or \acp{IQM} \cite{Chen_2019_OE, Chen_2019_AO, Zhang_2023_JLT}. Each \ac{SCE} contains a passive combiner with multiple output ports, e.g., a $120\degree$ optical hybrid as shown in Fig.~\ref{fig:OAWG_F2_SCE_setup}\,(a) or a $90\degree$ optical hybrid as shown in Fig.~\ref{fig:OAWG_F2_SCE_setup}\,(b). 
For the experiments discussed in Section~\ref{sec:OAWG_Exp} below, we relied on a $90\degree$ optical hybrid, which was readily available as a fiber-pigtailed component. The desired optical signal $\underline{a}_{\mathrm{out}}(t)$ is then obtained at one of the output ports of the passive combiner, marked yellow in Fig.~\ref{fig:OAWG_F2_SCE_setup}\,(a) and~(b), while two of the other ports are connected to a low-speed \ac{BPD} to generate an error signal $U_{\mathrm{err}}(t)$ that is used for feedback-based compensation of the underlying phase error $\Delta\varphi(t)$, see \textcolor{color_Supp1}{Supplement\,1} Section\,1 for a more detailed mathematical description. The error signal results from interference of the overlapping spectral components of neighboring tributaries, see marked \acfp{OR} in Insets~\symBCircRed{} and~\symCCircRed{} of Fig.~\ref{fig:OAWG_F1_Vision}. Under the assumption that the average optical powers of the combined tributaries within the overlap regions are constant and that the system is close to the desired operating point $\Delta\varphi = 0$ a linear approximation, which renders the error signal $U_{\mathrm{err}}(t)$ proportional to the phase error $\Delta\varphi(t)$, can be used, 
\begin{equation}
U_{\mathrm{err}}(t) \propto \mathrm{sin} \left( \Delta\varphi(t)\right) \approx  \Delta\varphi(t).
\label{eq:OAWG_EQ6_Uerr}
\end{equation}
The error signal $U_{\mathrm{err}}(t)$ is then fed to controller which drives a \acf{PS} at one of the optical input ports, Fig.~\ref{fig:OAWG_F2_SCE_setup}\,(a) and~(b). Note that depending on the insertion loss of each \ac{SCE} and the number of spectral slices combined, an optical amplifier at the output of the \ac{SCT} might be required to compensate for the unavoidable optical loss that is introduced by the rather simple wavelength-agnostic \ac{SCE} implementations shown in Fig.~\ref{fig:OAWG_F2_SCE_setup}\,(a) and~(b). In this respect, the $120\degree$ optical hybrid is preferred as it does not waste power to an unused port, denoted by “$\pi$” in Fig.~\ref{fig:OAWG_F2_SCE_setup}~(b), and since it can be easily implemented as a symmetrical 3\texttimes3 \ac{MMI} coupler. More advanced designs of \acp{SCE}, exploiting, e.g., wavelength-dependent \ac{PIC} elements such as arrayed waveguide gratings could mitigate optical losses further.

\begin{figure}[htb]
    \centering
    \includegraphics{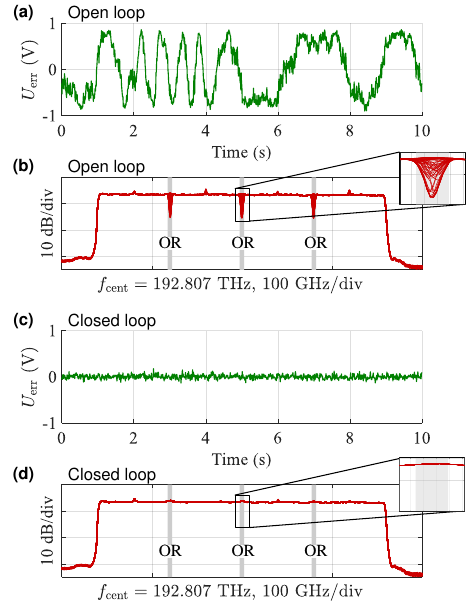}
    \caption{\acswitchoff Experimental verification of the actively phase-stabilizing \acf{SCT} based on $90\degree$ optical hybrids as passive combiners, see Fig.~\ref{fig:OAWG_F2_SCE_setup}\,(b) \subfigcap{(a)}~Exemplary error signal $U_{\mathrm{err}}(t)$ measured for an \acf{SCE} with deactivated control loop. Random variations of the error signal are observed as the phase of the different spectral slices drifts randomly. \subfigcap{(b)}~Overlay of 50 optical spectra recorded with deactivated control loop. The random drift of the relative phase $\Delta\varphi(t)$ between adjacent spectral slices leads to randomly occurring spectral dips due to constructive and destructive interference in the \aclp{OR} (\acsp{OR}, marked in gray). Note that the depth of the measured dips is limited due to the non-zero \acf{RBW} of the \acl{OSA} (\acs{OSA}, $\mathrm{RBW} = 2.48\,\mathrm{GHz}$). \subfigcap{(c)}~Residual error signal $U_{\mathrm{err}}(t)$ measured when the phase control is activated. \subfigcap{(f)}~Overlay of 50 optical spectra recorded with activated control loop. The relative phases of all spectral slices are aligned and stabilized, and no spectral dips can be found in the \acsp{OR} (marked in gray).}
    \label{fig:OAWG_F3_SCE_control_on_off}
\end{figure}

To verify the viability of the \ac{SCE}, we use fiber-optic components and an implementation based on a $90\degree$ optical hybrid, according to Fig.~\ref{fig:OAWG_F2_SCE_setup}\,(b). We record an exemplary error signal $U_{\mathrm{err}}(t)$ in the open-loop as well as in the closed-loop configuration, see Fig.~\ref{fig:OAWG_F3_SCE_control_on_off}\,(a) and~(c) respectively. As expected, using fiber optic components leads to a significant drift of the optical phase within ten seconds if the active phase stabilization is turned off. As a result, we observe randomly occurring destructive or constructive interference within the three \aclp{OR} of the overall output signal, leading to a random variation of the \ac{BPD}'s output voltage between approximately $-1\,\mathrm{V}$ and $+1\,\mathrm{V}$, Fig.~\ref{fig:OAWG_F3_SCE_control_on_off}\,(a). The randomly occurring destructive or constructive interference is visualized in Fig.~\ref{fig:OAWG_F3_SCE_control_on_off}\,(b), where 50 superimposed optical spectra of the overall output waveform $\underline{a}_{\mathrm{out}}(t)$ are shown and where strong random dips are observed in the three \aclp{OR}. By closing the control loop, the error signal $U_{\mathrm{err}}(t)$, Fig.~\ref{fig:OAWG_F3_SCE_control_on_off}\,(c), and consequently the phase error $\Delta\varphi(t)$ are minimized. As a result, all superimposed tributaries interfere constructively, and the spectral dips disappear as can be seen from the 50 superimposed spectra shown in Fig.~\ref{fig:OAWG_F3_SCE_control_on_off}\,(d). 
Note that our current experiments rely on a piezo-based fiber stretcher (FPS-003, General Photonics now part of Luna Innovations) with a tuning range of $55\pi$. When used with extended fiber-based setups, the fiber-stretcher typically reaches its limit within a few tens to a few 100s of seconds, which triggers an automatic reset of the underlying digital \ac{PI}-controller. This undesirable reset can be avoided in future implementations by using an endless phase shifter instead of the fiber-stretcher \cite{Ashok_2023_IEEE_JQE} or by integrating the system, such that a simple phase shifter with finite tuning range is sufficient \cite{Yamazaki_2021_JLT, Yamazaki_2023_JLT, Yamazaki_2024_JLT}.

 \begin{figure*}[htb]
        \centering       
        \includegraphics[width = \linewidth]{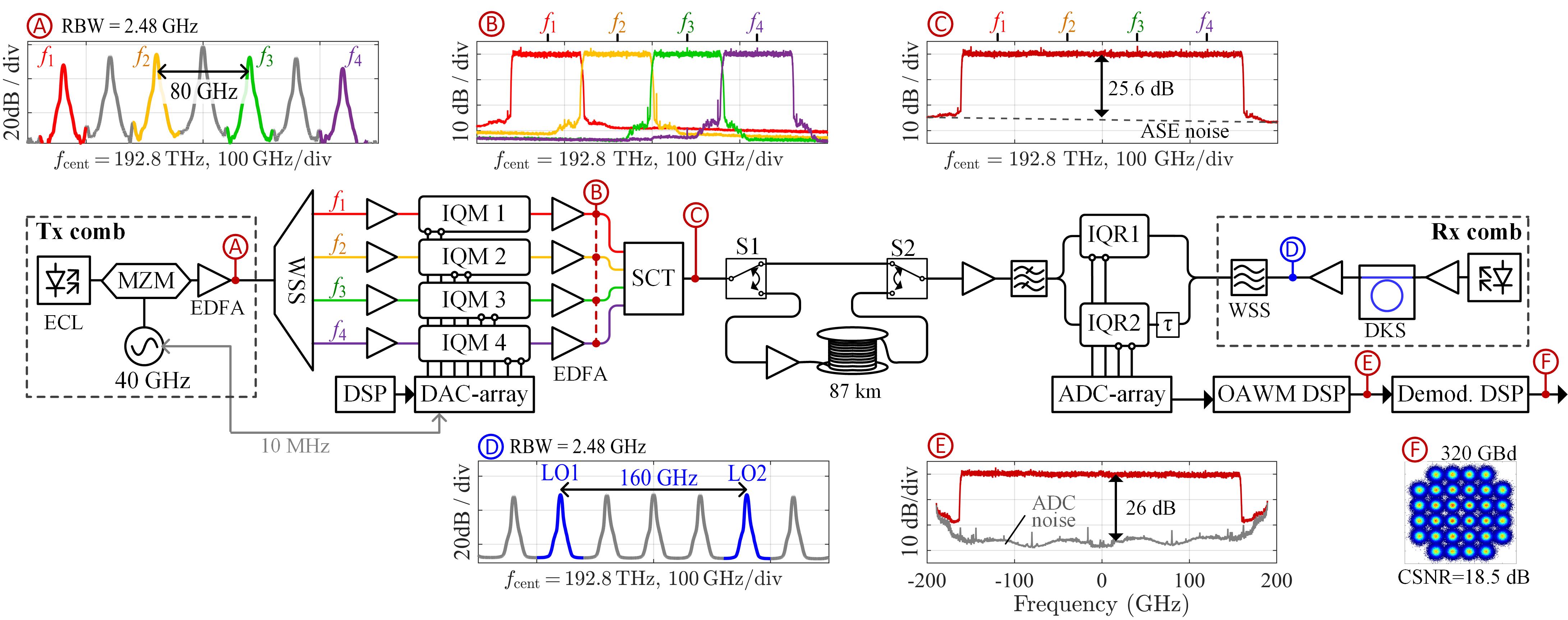}
        \caption{\acswitchoff Experimental setup of our \acs{OAWG} transmission experiment and exemplary measurement results taken at various points \symACircRed[0.9]{}...\symFCircRed[0.9]. The transmitter comb (Tx comb, Point~\symACircRed[0.9]{}) is generated by modulating a \acs{CW} tone emitted by an \acf{ECL}. The resulting Tx comb is amplified by an \acf{EDFA}, and individual tones $\color{color_slice_1_txt}f_1,\color{color_slice_2_txt}f_2,\color{color_slice_3_txt}f_3,\color{color_slice_4_txt}f_4$ are selected by a \acf{WSS} to serve as carriers for \acs{IQ} modulation. The drive signals for the \acs{IQ} modulators (\acs{IQM}1,..., \acs{IQM}4) are calculated by a \acf{DSP} block and generated by a \acs{DAC} array (Keysight M8194A) that is \acs{RF}-synchronized to the Tx comb generator. A phase-stabilizing \acf{SCT} combines all tributaries, Point~\symBCircRed[0.9]{}, thus forming the output waveform $\aSt$, Point~\symCCircRed[0.9]{}. The generated waveform is measured by a two-channel non-sliced \acs{OAWM} receiver -- either directly (upper position of switches S1 and S2) or after transmission through an 87\,km-long fiber link (lower switch positions). The \acs{OAWM} system uses two \acs{IQ} receivers (\acsp{IQR}) that are fed by the received waveform and by time-delayed copies of the Rx comb, comprising two tones (LO1, LO2, Inset~\symDCircBlue[0.9]{}) derived from a \acf{DKS} comb. The photocurrents of the \acsp{IQR} are digitized by an \acs{ADC} array (Keysight UXR series oscilloscope), and are used to reconstruct the received waveform via the \acs{OAWM} \acs{DSP}, Point~\symECircRed[0.9]{}. The reconstructed waveform is then demodulated (Demod. DSP) to retrieve the transmitted data, Point~\symFCircRed[0.9]{}. Inset~\symACircRed[0.9]{}:\,Optical spectrum of Tx comb. Note that the displayed spectral width of the individual comb-tones is dictated by the rather large \acf{RBW} of the spectrum analyzer that was used for the measurement (\acs{RBW}\,=\,2.48\,GHz). Inset~\symBCircRed[0.9]{}:\,High-resolution (\acs{RBW}\,=\,100\,MHz) optical spectra of individual the tributary signals. Inset~\symCCircRed[0.9]{}:\,High-resolution (\acs{RBW}\,=\,100\,MHz) optical spectrum of 320\,GBd 16QAM signal. Inset~\symDCircBlue[0.9]{}:\,Optical spectrum of the Rx comb (\acs{RBW}\,=\,2.48\,GHz) that is used for \acs{OAWM}. Inset~\symECircRed[0.9]{}: Spectrum of reconstructed 320\,GBd 16QAM waveform obtained from the \acs{OAWM} receiver in the optical back-to-back configuration (\acs{RBW}\,=\,100\,MHz). Inset~\symFCircRed[0.9]{}:\,Constellation diagram and \acf{CSNR} for an exemplary 320\,GBd 32QAM signal measured in back-to-back configuration .}
        \label{fig:OAWG_F4_exp_setup}
\end{figure*}

\section{320 GBd transmission experiment}
\label{sec:OAWG_Exp}

\subsection{Experimental setup }
\label{sec:OAWG_Exp_setup}

To demonstrate the viability of the scheme described in Section~\ref{sec:OAWG_VisionAndConcept}, we perform a proof-of-concept experiment in which we transmit and receive broadband optical communication signals with symbol rates of up to 320\,GBd by combining phase-stabilized \ac{OAWG} with non-sliced \ac{OAWM}, see Fig.~\ref{fig:OAWG_F4_exp_setup} for the associated experimental setup. The \ac{OAWG} subsystem relies on a transmitter frequency comb (Tx comb) which is generated by modulating a \ac{CW} laser tone using \iac{MZM} which is driven by a 40\,GHz sinusoidal. The Tx comb is subsequently amplified by an \ac{EDFA}, and four phase-locked tones at frequencies $f_1, f_2, f_3$, and $f_4$ are selected by a \ac{WSS} to serve as carriers for spectrally sliced signal synthesis. We measure the spectrum of the Tx comb at point \symACircRed{} of Fig.~\ref{fig:OAWG_F4_exp_setup}, see corresponding inset. The \ac{OCNR} of all comb lines exceeds 30\,dB, measured with respect to the standard reference bandwidth of 12.5\,GHz, which corresponds to a wavelength interval of 0.1\,nm at a center wavelength of $\lambda = 1550$\,nm. The isolated optical carriers are amplified and fed to an array of four \ac{IQ}-modulators (\acs{IQM}\,1 $\ldots$ \acs{IQM}\,4), which are electrically driven by an overall eight \ac{DAC} outputs (\ac{DAC} array) of two \aclp{AWG} (Keysight M8194A), synchronized to the Tx-comb via a 10\,MHz reference clock. The output signals of the \acp{IQM} are amplified (point \symBCircRed{} in Fig.~\ref{fig:OAWG_F4_exp_setup}) and combined by an \ac{SCT} that ensures stable phase relationships among all superimposed tributaries.

The electrical signals $\Inut$ and $\Qnut$ driving the various \acp{IQM} are pre-distorted based on Eq.~\ref{eq:OAWG_EQ5_system_FWRD_matrix}, using the measured transfer functions $\HnuIf$ and $\HnuQf$ that comprise the characteristics of the \ac{DAC} array, the \acp{IQM}, and the electrical and optical amplifiers. The drive signals are clipped to a \ac{PAPR} of 10\,dB to reduce the impact of quantization noise of the \acp{DAC}. Insets~\symBCircRed{} and~\symCCircRed{} of Fig.~\ref{fig:OAWG_F4_exp_setup} depict the optical power spectra of the four tributaries and of the associated merged output waveform, for a 320\,GBd 16\acs{QAM} signal. We use \ac{RRC} pulse shapes with a roll-off of $\rho = 0.01$, leading to a spectrally flat power spectrum, which is nicely reproduced by the directly measured optical spectrum in Insets~\symBCircRed{} and~\symCCircRed{} of Fig.~\ref{fig:OAWG_F4_exp_setup}, thus confirming the viability of our signal-generation and pre-distortion scheme. From the spectrum shown in Inset~\symCCircRed{} of Fig.~\ref{fig:OAWG_F4_exp_setup}, we estimate the in-band \ac{ASE} noise level caused by the various \acp{EDFA} by interpolating the out-of-band noise, indicated by a dashed line in Inset~\symCCircRed{} of Fig.~\ref{fig:OAWG_F4_exp_setup}. Note that the overall in-band noise exceeds the pure \ac{ASE} noise as it additionally comprises \ac{DAC} noise, noise from electrical amplifiers, as well as distortions such as remaining \ac{IQ} imbalance. We measure the overall transmitter noise and distortions by generating single-sideband signals and estimate the \ac{SNDR} to be in the range of 20\,dB for a 320\,GBd optical waveform, see \textcolor{color_Supp1}{Supplement\,1} Section\,2.2 for details. 

\begin{figure*}[htb] 
\centering
\input{Abbildungen/OAWG_F5_SyR_sweep_data_comparison_tikZ}
\caption{\acswitchoff Generation and measurement of \acf{QAM} signals with symbol rates of up to 320\,GBd using OAWG and OAWM techniques, and comparison to competing approaches. \subfigcap{(a)}Optical spectra of 16QAM signals with symbol rates ranging from 80\,GBd to 320\,GBd generated using spectrally sliced \acs{OAWG} and measured using a high-resolution \acl{OSA} (\acs{OSA}, Apex AP2060, resolution bandwidth 100 MHz). For better comparison, all spectra are normalized to the out-of-band \acf{ASE} noise level. We indicate the frequencies $f_1, f_2, f_3$ and $f_4$ of the Tx comb, for which a constant \acf{FSR} is maintained throughout the experiment. \subfigcap{(b)}\Acf{CSNR} as a function of the symbol rate for 16QAM (blue dots and solid blue line) and 32QAM signals (red dots and red dashed line) -- measured in an optical back-to-back (ob2b) configuration and after transmission over 87\,km of single-mode fiber (blue and red cross, see inset). The results are compared to other high-symbol-rate optical signaling experiments that rely on single \acfp{DAC} \cite{Chen_2022_OFC, Zheng_2023_JLT, Xu_2022_Optica}, purely electronic \cite{Nakamura_2019_ECOC, Nakamura_2019_OFC, Nakamura_2019_OFC_2, Nagatani_2020_IEEE_JSSC, Heni_2020_JLT, Pittala_2020_ECOC, Mardoyan_2022_OFC, Hu_2022_JLT, Chen_2020_ECOC, Chen_2021_OFC, Mardoyan_2022_ECOC, Almonacil_2023_JLT, Nakamura_2022_ECOC, Berikaa_2023_PTL, Ozaki_2023_JLT} (square markers), or photonic-electronic \cite{RiosMuller_2015_OFC, Yamazaki_2021_JLT, Henauer_2022_OFC} (circular markers) multiplexing techniques. References \cite{Hu_2022_JLT, Heni_2020_JLT, Berikaa_2023_PTL} demonstrate \acf{PAM} signaling, whereas the other publications show \acs{QAM} signals. References \cite{Pittala_2020_ECOC, Mardoyan_2022_OFC, Hu_2022_JLT} and \cite{Mardoyan_2022_ECOC,  Almonacil_2023_JLT, Nakamura_2022_ECOC, Berikaa_2023_PTL} use the commercially available signal generators Keysight M8199A and M8199B, respectively, which rely two time-interleaved \acs{DAC} channels.\quad Insets:\,Exemplary constellation diagrams for 16QAM and 32QAM 320\,GBd signals and measured \acf{BER} obtained for the optical back-to-back configuration.}
\label{fig:OAWG_F5_SyR_sweep_tikZ}
\end{figure*}

The generated optical signal is then either fed directly to the \ac{OAWM} receiver in an optical back-to-back configuration, corresponding to the upper position of switches S1 and S2 in Fig.~\ref{fig:OAWG_F4_exp_setup}, or it is first sent through a booster amplifier and transmitted over an 87\,km-long fiber link associated with the lower switch positions. At the receiver, we rely on a two-channel non-sliced \ac{OAWM} system, see \cite{Drayss_2023_Optica} for details on the underlying receiver concept. The \ac{OAWM} receiver uses a free-running \acf{DKS} comb (Rx comb) as a multi-wavelength \ac{LO} \cite{Kippenberg_2018_Science}, from which we extract two phase-locked tones spaced by approximately 160\,GHz by means of a second \ac{WSS}. The spectrum of the Rx comb is depicted in Inset~\symDCircBlue{} of Fig.~\ref{fig:OAWG_F4_exp_setup}. Both comb lines have an \ac{OCNR} of 30\,dB with respect to the standard reference bandwidth of 12.5\,GHz. For high-fidelity signal reconstruction, we calibrate the \ac{OAWM} receiver using a known optical reference waveform generated by a femtosecond mode-locked laser, see \cite{Drayss_2023_Optica, Drayss_2024_JLT} for details. The power spectrum of the reconstructed waveform measured at point~\symECircRed{} of Fig.~\ref{fig:OAWG_F4_exp_setup} is shown in the corresponding inset. We additionally include the power spectral density of \ac{ADC} noise, measured after applying the \ac{OAWM} \ac{DSP} (gray trace, Inset~\symECircRed{} of Fig.~\ref{fig:OAWG_F4_exp_setup}), which is approximately 26\,dB below the signal -- comparable to the impairments by Tx \ac{ASE} noise. From the waveform reconstructed by the \ac{OAWM} receiver, we finally recover the constellation diagram of the 320\,GBd 32QAM data signal shown in Inset~\symFCircRed{} of Fig.~\ref{fig:OAWG_F4_exp_setup}, which leads to a \ac{CSNR} of $\mathrm{CSNR}_{\mathrm{dB}} = 18.5\,\mathrm{dB}$. The \ac{CSNR} considers not only noise but also linear and nonlinear distortions and corresponds to the square of the reciprocal of the \ac{EVM} normalized to the average signal power \cite{Schmogrow_2012_PTL}, $\mathrm{CSNR}_{\mathrm{dB}}=10\mathrm{log}_{10}\bigl(1/\mathrm{EVM}_{\mathrm{a}}^2\bigr)$, see  Section\,3 of \textcolor{color_Supp1}{Supplement\,1} for details.

\subsection{Experimental results and discussion}
\label{sec:OAWG_Exp_results}

We use the system described in Section~\ref{sec:OAWG_Exp_setup} and Fig.~\ref{fig:OAWG_F4_exp_setup} to generate and receive 16QAM and 32QAM waveforms with symbol rates ranging from 80\,GBd to 320\,GBd -- either in an optical back-to-back configuration or after propagating through the 87\,km fiber link. Figure~\ref{fig:OAWG_F5_SyR_sweep_tikZ}\,(a) shows the synthesized spectra of data signals of various symbol rates measured using a high-resolution \acl{OSA} (\acs{OSA}, Apex AP2060, \acs{RBW}\,=\,100\,MHz) directly at the \ac{OAWG} transmitter output, i.e., at Point~\symCCircRed{} in Fig.~\ref{fig:OAWG_F4_exp_setup}. The frequencies $f_1,...,f_4$ of the four optical carriers are also indicated in Fig.~\ref{fig:OAWG_F5_SyR_sweep_tikZ}\,(a). Note that, due to limitations of the underlying \ac{RF} components, the \ac{FSR} of the Tx comb was kept constant throughout the experiment such that lower symbol rates resulted in only partially filled or even empty spectral slices. Figure~\ref{fig:OAWG_F5_SyR_sweep_tikZ}\,(b) shows the \ac{CSNR} obtained for various symbol rates measured in the optical back-to-back configuration for 16QAM signals (blue dots, solid line) and for 32QAM signals (red dots, dashed line). For the highest symbol rate of 320\,GBd both the 16QAM and the 32QAM signals were sent over the 87\,km fiber link. After applying digital dispersion compensation, we find a \ac{CSNR} penalty is 0.5\,dB compared to the optical back-to-back configuration, see red and blue cross on the right of Fig.~\ref{fig:OAWG_F5_SyR_sweep_tikZ}\,(b) and in the corresponding inset. Compared to the most advanced competing high symbol-rate \acs{QAM} and \acs{PAM} signaling experiments \cite{RiosMuller_2015_OFC, Yamazaki_2021_JLT, Henauer_2022_OFC, Nakamura_2019_ECOC, Nakamura_2019_OFC, Nakamura_2019_OFC_2, Nagatani_2020_IEEE_JSSC, Heni_2020_JLT, Pittala_2020_ECOC, Mardoyan_2022_OFC, Xu_2022_Optica, Hu_2022_JLT, Chen_2020_ECOC, Chen_2021_OFC, Chen_2022_OFC, Zheng_2023_JLT, Mardoyan_2022_ECOC, Almonacil_2023_JLT, Nakamura_2022_ECOC, Berikaa_2023_PTL, Ozaki_2023_JLT} relying on optical  \cite{RiosMuller_2015_OFC, Yamazaki_2021_JLT, Henauer_2022_OFC}, or electrical \cite{Nakamura_2019_ECOC, Nakamura_2019_OFC, Nakamura_2019_OFC_2, Nagatani_2020_IEEE_JSSC, Heni_2020_JLT, Pittala_2020_ECOC, Mardoyan_2022_OFC, Hu_2022_JLT, Chen_2020_ECOC, Chen_2021_OFC, Mardoyan_2022_ECOC, Almonacil_2023_JLT, Nakamura_2022_ECOC, Berikaa_2023_PTL, Ozaki_2023_JLT} multiplexing techniques, our experiments show significant advantages regarding the symbol rates and the \ac{CSNR}, see additional data points and corresponding references in Fig.~\ref{fig:OAWG_F5_SyR_sweep_tikZ}\,(b). Specifically, as the spectral multiplexing is done in the optical domain, our approach does not suffer from performance degradation for symbol rates above 200\,GBd, as typically observed for the purely electronic approaches \cite{Nakamura_2019_ECOC,Nakamura_2019_OFC, Nakamura_2019_OFC_2, Nagatani_2020_IEEE_JSSC, Heni_2020_JLT, Pittala_2020_ECOC, Mardoyan_2022_OFC, Xu_2022_Optica,Hu_2022_JLT, Chen_2020_ECOC, Chen_2021_OFC, Chen_2022_OFC, Zheng_2023_JLT, Mardoyan_2022_ECOC,Almonacil_2023_JLT,Nakamura_2022_ECOC,Berikaa_2023_PTL,Ozaki_2023_JLT}. The OAWG approach hence renders the achievable optical bandwidth independent of the bandwidth of the underlying electronic components. Note also that the use of demultiplexing filters for separating the frequency comb tones in combination with the active phase stabilization and the accurate transmitter calibration avoids the formation of spectral images that have been observed in other optical multiplexing techniques that, e.g., rely on optical time or phase interleaving \cite{Yamazaki_2021_JLT, Yamazaki_2023_JLT}.

Our \ac{OAWG} transmitter and the associated \ac{OAWM} receiver hence offer attractive performance advantages not only in terms of the achievable bandwidth but also in terms of the overall signal quality, while leaving room for further improvement. The dominant noise sources in our system are \ac{ASE} noise of the various \acp{EDFA}, \ac{DAC} noise, \ac{ADC} noise, and the limited \ac{OCNR} of the Tx and Rx comb. Further \ac{SNDR} improvements are possible, e.g., by adding optical bandpass filters after each \ac{IQM} to suppress \ac{ASE} noise or by pre-emphasizing the high-frequency components via a programmable optical filter that compensates the \ac{RF} frequency responses of the \ac{DAC}-array and the \acp{IQM} in the optical domain \cite{Xu_2022_Optica, Mardoyan_2022_ECOC, Almonacil_2023_JLT, Nakamura_2022_ECOC}. The receiver noise could be further reduced by increasing the channel count of the \ac{OAWM} system and by relying on a larger number of lower-speed \ac{ADC}, see ref. \cite{Drayss_2024_JLT} for a more detailed discussion. Similarly, the \ac{OAWG} performance could be improved by using more \ac{DAC} channels or by adapting the \ac{FSR} of the Tx comb to the symbol rate. Specifically, for a fixed \ac{FSR} as in our current implementation, reduced symbol rates lead to only partially filled or even empty spectral slices, see Fig.~\ref{fig:OAWG_F5_SyR_sweep_tikZ}\,(a), while all active \ac{DAC} channels and associated \acp{IQM} still run at the maximum \ac{RF} bandwidth of ~40\,GHz. This leads to noticeable impairments through bandwidth limitations and limited \ac{SNDR} of the various drive signals, which could be mitigated by using spectral slices with a smaller uniform bandwidth.

Still, the signals generated and measured using the setup in Fig.~\ref{fig:OAWG_F4_exp_setup} already show excellent signal quality, Fig.~\ref{fig:OAWG_F5_SyR_sweep_tikZ}\,(b). To the best of our knowledge, the 320\,GBd demonstrated in our transmission experiments represent the highest symbol rate so far achieved for “fully coherent” \ac{QAM} data signals that do not rely on \ac{OTDM}-based schemes, for which subsequent symbols do not have a fixed phase relationship and where the pulse shape is defined by a mode-locked laser rather than synthesized digitally \cite{ThomasRichter_2012_JLT}. We believe that our work can go beyond high-bandwidth communications and pave a way towards photonic-electronic \acp{DAC} with unprecedented bandwidth \cite{Fullner_2022_CLEO}, terahertz signal generation in future wireless networks \cite{Harter_2018_NatPhoton,Harter_2019_Optica}, fully software-defined optical transceivers \cite{YanniOu_2016_JLT}. or advanced test and measurement equipment.

\section{Summary}
\label{sec:Summary}

We have demonstrated the generation and transmission of fully coherent 16QAM and 32QAM signals at record-high symbol rates up to 320\;GBd by combining comb-based \ac{OAWG} and \ac{OAWM}. To the best of our knowledge, our work represents the first \ac{OAWG} demonstration using actively phase-stabilized spectral slices, leading to targeted optical waveform synthesis at the highest bandwidth so far achieved in any \ac{OAWG} experiment. By multiplexing parallel \ac{ADC} and \ac{DAC} arrays in the optical domain, our concept renders the bandwidth of the optical signal independent of the bandwidth of individual electronic interfaces. This may offer a path forward towards ultra-broadband photonic-electronic signal generators that are relevant in a variety of technical and scientific applications.

\begin{backmatter}
\bmsection{Acknowledgments} This work was supported by the ERC Consolidator Grant TeraSHAPE (\# 773248) and by the associated ERC Proof-of-Concept Grant TeraGear (\# 101123567), by the EU H2020 project TeraSlice (\# 863322), by the DFG projects PACE (\# 403188360) and GOSPEL (\# 403187440), by the joint DFG-ANR projects HybridCombs (\# 491234846) and Quad- Combs (\# 505515860 ), by the DFG Collaborative Research Centers (CRC) WavePhenomena (SFB 1173, \# 258734477) and HyPERION (SFB 1527,
\# 454252029), by the BMBF project Open6GHub (\# 16KISK010), by the Horizon Europe EIC transition program with the projects CombTools (\# 101136978), MAGNIFY (\# 101113302), and HDLN (\#101113260), by the Alfried Krupp von Bohlen und Halbach Foundation, by the MaxPlanck School of Photonics (MPSP), and by the Karlsruhe School of Optics \& Photonics (KSOP).

\end{backmatter}
\bibliography{manuscript}
\bibliographyfullrefs{manuscript}

\clearpage



\section*{Supplementary material}

# Optical Arbitrary Waveform Generation (OAWG) Using Actively Phase-Stabilized Spectral Stitching: Supplementary document

Daniel Drayss[1,2,+,*], Dengyang Fang[1,+], Alban Sherifaj[1,+], Huanfa Peng[1], Christoph Füllner[1], Thomas Henauer[3], Grigory Lihachev[4], Tobias Harter[1], Wolfgang Freude[1], Sebastian Randel[1], Tobias J. Kippenberg[4], Thomas Zwick[3], and Christian Koos[1,2,**]

[1] *Institute of Photonics and Quantum Electronics (IPQ), Karlsruhe Institute of Technology (KIT), 76131 Karlsruhe, Germany*
[2] *Institute of Microstructure Technology (IMT), Karlsruhe Institute of Technology (KIT), 76344 Eggenstein-Leopoldshafen, Germany*
[3] *Institute of Radio Frequency Engineering and Electronics (IHE), Karlsruhe Institute of Technology (KIT), 76131 Karlsruhe, Germany*
[4] *Institute of Physics, Swiss Federal Institute of Technology Lausanne (EPFL), CH-1015 Lausanne, Switzerland*
[+] *Contributed equally.*    [*] *daniel.drayss@kit.edu.*    [**] *christian.koos@kit.edu*

This document provides supplementary information to the research paper entitled "Optical Arbitrary Waveform Generation (OAWG) Using Actively Phase-Stabilized Spectral Stitching:". In Section 1, we discuss the generation of the error signal for the feedback-controlled phase stabilization. The system calibration techniques are explained in Section 2, where we also estimate the signal-to-noise-and-distortion ratio (SNDR) achieved by the OAWG system. Section 3 provides a definition of the constellation signal-to-noise ratio (CSNR), which is used in the main manuscript to quantize the signal quality.

## 1. Mathematical derivation of error-signal generation for closed loop phase control

In the main manuscript, we stated that the phase error $\Delta\varphi(t)$ between two overlapping tributary signals, e.g., $\underline{a}_{\text{in},1}(t)$ and $\underline{a}_{\text{in},2}(t)\exp(j\Delta\varphi(t))$ as in Fig. S1 (a) below, can be measured by using a suitable passive combiner in conjunction with a low-speed balanced photodetector (BPD). In the following, we provide a more detailed derivation of the underlying mathematical relations for either a 120° optical hybrid or a 90° optical hybrid as a passive combining element. For the experiments discussed in the main manuscript, we have used a 90° optical hybrid, which is a standard and readily available component in coherent optical communications. The digital proportional-integral (PI) controller was implemented on an open-source software measurement and control board (Red Pitaya STEMlab 125 with Xilinx Zynq 7010) and a piezo-based fiber stretcher (FPS-003, General Photonics now part of Luna Innovations) with a tuning range of $55\pi$ was used as a phase shifter (PS).

### 1.1. Error signal generation using 120°optical hybrid

Figure S1 (a) shows a signal-combining element (SCE) using a 120°optical hybrid as a passive combiner, which may, e.g., be implemented based on a 3×3 multi-mode interference (MMI) coupler. All output signals and components related to the generation of the error signal $U_{\text{err}}(t)$ are marked in green. For simplicity, the signal tributaries $\underline{a}_{\text{in},1}(t)$ and $\underline{a}_{\text{in},2}(t)$ are assumed to spectrally overlap and to contain identical signal components within the overlap region (OR), $[f_2 - B, f_1 + B]$, see gray area in Fig. S1 (b). The signal components of the two tributaries in the overlap region are referred to by $\underline{a}_{\text{OR}}(t)$, where the corresponding spectrum $\underline{\tilde{a}}_{\text{OR}}(f)$ is given by

$$\underline{\tilde{a}}_{\text{OR}}(f) = \begin{cases} \underline{\tilde{a}}_{\text{in},1}(f) = \underline{\tilde{a}}_{\text{in},2}(f) & \text{for } f \in [f_2 - B, f_1 + B] \\ 0 & \text{otherwise.} \end{cases} \tag{S1}$$

To derive a mathematical model of the SCE, we start from the scattering matrix of an ideal 3×3 MMI [1], where the center input port is unused, Fig. S1 (a),

$$\begin{bmatrix} \underline{a}_{\text{cntrl},1}(t) \\ \underline{a}_{\text{out}}(t) \\ \underline{a}_{\text{cntrl},2}(t) \end{bmatrix} = \frac{1}{\sqrt{3}} \begin{bmatrix} -1 & e^{j\pi/3} \\ 1 & 1 \\ e^{j\pi/3} & -1 \end{bmatrix} \begin{bmatrix} \underline{a}_{\text{in},1}(t) \\ \underline{a}_{\text{in},2}(t)e^{j\Delta\varphi(t)} \end{bmatrix} \tag{S2}$$

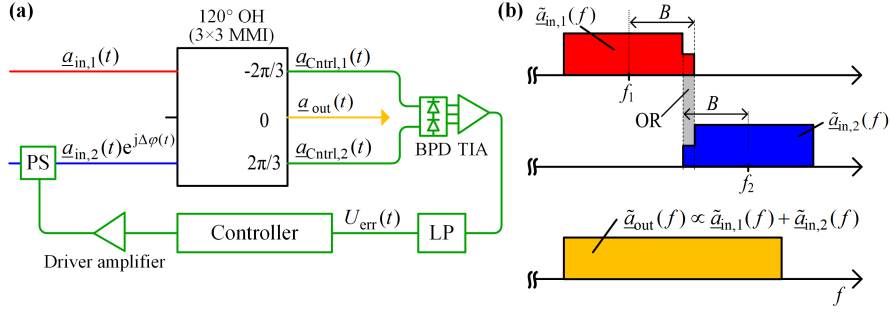

**Fig. S1.** Actively phase-stabilized combining of two overlapping spectral slices using a signal-combining element (SCE) based on a 120° optical hybrid (OH). **(a)** Conceptual setup of an SCE using a 3×3 multi-mode interference (MMI) coupler, for which the center input port remains unused. The MMI coupler superimposes the input signals $\underline{a}_{\text{in},1}(t)$ and $\underline{a}_{\text{in},2}(t)\exp(\text{j}\Delta\varphi(t))$, having a phase error $\Delta\varphi(t)$ with relative phases of $-2\pi/3$, $0$, and $2\pi/3$ at the various output ports. The outer output ports (green) are used to generate the optical control signals $\underline{a}_{\text{cntrl},1}(t)$ and $\underline{a}_{\text{cntrl},2}(t)$, whereas the combined output signal $\underline{a}_{\text{out}}(t)$ is generated at the center "zero phase" port ("0", yellow). The electrical error signal $U_{\text{err}}(t)$ is obtained by feeding the two optical control signals $\underline{a}_{\text{cntrl},1}(t)$ and $\underline{a}_{\text{ctnrl},2}(t)$ to a low-speed balanced photodetector (BPD), which is followed by a transimpedance amplifier (TIA) and a low-pass filter (LP). In our model, the LP represents the bandwidth limitations of the entire loop, including the BPD, the TIA, the controller, the driver amplifier, and the phase shifter (PS). The low-pass filtered error signal $U_{\text{err}}(t)$ is essentially proportional to phase error $\Delta\varphi(t)$, using a linear approximation close to the desired operating point $\Delta\varphi = 0$, see Eq. S10. A proportional-integral (PI) controller can hence be used to drive the phase shifter (PS) and to compensate for the measured phase error $\Delta\varphi(t)$. **(b)** Spectra $\underline{\tilde{a}}_{\text{in},1}(f)$ and $\underline{\tilde{a}}_{\text{in},2}(f)$ of the two tributaries $\underline{a}_{\text{in},1}(t)$ and $\underline{a}_{\text{in},2}(t)$ along with the spectrum of the targeted output waveform $\underline{a}_{\text{out}}(t) \propto \underline{a}_{\text{in},1}(t) + \underline{a}_{\text{in},2}(t)$ for $\Delta\varphi(t) = 0$. The spectra of the two tributaries are overlapping within the overlap region (OR, $f \in [f_2 - B, f_1 + B]$, gray).

The output signal $\underline{a}_{\text{out}}(t)$ of the SCE is tapped at the center output port, labeled by a relative phase of "0" in Fig. S1 (a) and thus contains the sum of the two optical input signals $\underline{a}_{\text{in},1}(t)$ and $\underline{a}_{\text{in},2}(t)e^{\text{j}\Delta\varphi(t)}$,

$$\underline{a}_{\text{out}}(t) = \frac{1}{\sqrt{3}} \left[ \underline{a}_{\text{in},1}(t) + \underline{a}_{\text{in},2}(t)e^{\text{j}\Delta\varphi(t)} \right] \tag{S3}$$

To obtain the targeted output signal, the slowly varying phase error $\Delta\varphi(t)$ between the two tributaries at the input of the MMI must be kept at zero by the feedback loop and the phase shifter (PS). The error signal required for the control loop is generated by balanced detection of the two optical signals $\underline{a}_{\text{cntrl},1}(t)$ and $\underline{a}_{\text{cntrl},2}(t)$, which are obtained at the upper and the lower MMI output labeled by relative phases $-2\pi/3$ and $2\pi/3$ in Fig. S1 (a) and by subsequent low-pass filtering,

$$U_{\text{err}}(t) = H_{\text{LP}}(t) * \left[ |\underline{a}_{\text{cntrl},1}(t)|^2 - |\underline{a}_{\text{cntrl},2}(t)|^2 \right]. \tag{S4}$$

In this relation, $H_{\text{LP}}(t)$ is the impulse response of an equivalent low-pass filter, which represents the bandwidth limitations of the entire loop, including not only the BPD and the transimpedance amplifier (TIA) but also the lowpass characteristics of the subsequent components such as the controller, the driver amplifier and the phase shifter, Fig. S1 (a). The bandwidth $B_{\text{LP}}$ of the lowpass filter is chosen big enough to enable tracking of phase fluctuations on a millisecond time scale, but not too large as to suppress noise in the feedback loop. In our experiment, a bandwidth $B_{\text{LP}} \approx 10\,\text{kHz}$ turned out to be a good choice. Inserting Eq. S2 into Eq. S4 leads to

$$\begin{aligned} U_{\text{err}}(t) &= H_{\text{LP}}(t) * \left[ -\frac{4}{3}\sin(2\pi/3)\,\Im\left\{ \underline{a}_{\text{in},1}(t)\underline{a}_{\text{in},2}^*(t)e^{-\text{j}\Delta\varphi(t)} \right\} \right] \\ &= H_{\text{LP}}(t) * \left[ -\frac{2}{\sqrt{3}}\,\Im\left\{ \underline{a}_{\text{in},1}(t)\underline{a}_{\text{in},2}^*(t)e^{-\text{j}\Delta\varphi(t)} \right\} \right], \end{aligned} \tag{S5}$$

where $\Im\{\cdot\}$ refers to the imaginary part of the respective complex number. Assuming that the bandwidth $B_{\text{LP}}$ of the lowpass filter is sufficiently broad to capture the phase drift $\Delta\varphi(t)$ of our setup without distortions, i.e., $\exp\{-\text{j}\Delta\varphi(t)\}$ is approximately constant within the duration of the low-pass impulse response $H_{\text{LP}}(t)$,

2

we can simplify Eq. S5 by moving the phase-error term $\exp\{-\mathrm{j}\Delta\varphi(t)\}$ out of the convolution integral,

$$\begin{aligned}
U_{\mathrm{err}}(t) &\propto \Im\left\{H_{\mathrm{LP}}(t) * \left[\underline{a}_{\mathrm{in},1}(t)\underline{a}^*_{\mathrm{in},2}(t)\mathrm{e}^{-\mathrm{j}\Delta\varphi(t)}\right]\right\} \\
&= \Im\left\{\int_{-\infty}^{\infty} H_{\mathrm{LP}}(\tau)\underline{a}_{\mathrm{in},1}(t-\tau)\underline{a}^*_{\mathrm{in},2}(t-\tau)\mathrm{e}^{-\mathrm{j}\Delta\varphi(t-\tau)}\mathrm{d}\tau\right\} \\
&\approx \Im\left\{\mathrm{e}^{-\mathrm{j}\Delta\varphi(t)}\int_{-\infty}^{\infty} H_{\mathrm{LP}}(\tau)\underline{a}_{\mathrm{in},1}(t-\tau)\underline{a}^*_{\mathrm{in},2}(t-\tau)\mathrm{d}\tau\right\} \\
&= \Im\left\{\mathrm{e}^{-\mathrm{j}\Delta\varphi(t)}U_0(t)\right\},
\end{aligned} \tag{S6}$$

In this relation, the control-voltage amplitude $U_0(t)$ corresponds to the low-pass-filtered interference of the phase-stabilized overlapping signal components $\underline{a}_{\mathrm{in},1}(t)$ and $\underline{a}_{\mathrm{in},2}(t)$, while the phase error is contained in the pre-factor $\exp\{-\mathrm{j}\Delta\varphi(t)\}$.

$$U_0(t) = \int_{-\infty}^{\infty} H_{\mathrm{LP}}(t-\tau)\underline{a}_{\mathrm{in},1}(\tau)\underline{a}^*_{\mathrm{in},2}(\tau)\mathrm{d}\tau. \tag{S7}$$

Note that the low-pass filter bandwidth $B_{\mathrm{LP}}$ is much smaller than the bandwidth $B_{\mathrm{OR}}$ of the overlap region $B_{\mathrm{OR}}$, $B_{\mathrm{OR}}/B_{\mathrm{LP}} > 10^5$ in case of our experiments. As a consequence, for a given spectral component $\tilde{\underline{a}}_{\mathrm{in},1}(f_{\mathrm{a}})$ of the first tributary, the control-voltage amplitude $U_0(t)$ only contains beat signals with directly adjacent spectral components $\tilde{\underline{a}}_{\mathrm{in},2}(f_{\mathrm{b}})$, $|f_{\mathrm{a}} - f_{\mathrm{b}}| < B_{\mathrm{LP}}$, of the second tributary. The contribution of $\tilde{\underline{a}}_{\mathrm{in},1}(f)$ and $\tilde{\underline{a}}_{\mathrm{in},2}(f)$ to the control-voltage amplitude $U_0(t)$ in Eq. S7 is hence essentially limited to spectral regions, in which both $\tilde{\underline{a}}_{\mathrm{in},1}(f)$ and $\tilde{\underline{a}}_{\mathrm{in},2}(f)$ are nonzero, i.e., the overlap region $f \in$ OR. The tributaries $\underline{a}_{\mathrm{in},1}(t)$ and $\underline{a}_{\mathrm{in},2}(t)$ in Eq. S7 can thus be replaced by the common signal component $\underline{a}_{\mathrm{OR}}(t)$ within the overlap region, Eq. S1,

$$U_0(t) \approx H_{\mathrm{LP}}(t) * |\underline{a}_{\mathrm{OR}}(t)|^2, \qquad \text{for } B_{\mathrm{LP}} \ll B_{\mathrm{OR}}. \tag{S8}$$

We further make use of the fact that communication signals as used in our experiments are stationary and that both the symbol rate and the bandwidth of the overlap region greatly exceed the bandwidth $B_{\mathrm{LP}}$ of the low-pass filter. The low-pass filter hence averages over many symbols and we can assume that the resulting mean power $P_{\mathrm{OR}}$ of the signal components in the overlap region is constant, $P_{\mathrm{OR}} \propto H_{\mathrm{LP}}(t) * |\underline{a}_{\mathrm{OR}}(t)|^2 \approx$ const. The error signal in Eq. S6 can thus be re-written by replacing $U_0(t)$ by a constant $U_0$,

$$U_{\mathrm{err}}(t) \propto U_0 \Im\left\{\mathrm{e}^{-\mathrm{j}\Delta\varphi(t)}\right\} = U_0 \sin\{\Delta\varphi(t)\}. \tag{S9}$$

Once the feedback-based phase control is activate, the residual phase error $\Delta\varphi(t)$ is small, $\Delta\varphi(t) \ll 1$. We can thus linearize Eq. S9 about the desired operating point $\Delta\varphi = 0$ of the control loop,

$$U_{\mathrm{err}}(t) \propto \sin\{\Delta\varphi(t)\} \approx \Delta\varphi(t), \qquad \text{for } |\Delta\varphi| \ll 1. \tag{S10}$$

The error signal $U_{\mathrm{err}}(t)$ is hence approximately proportional to the phase error $\Delta\varphi(t)$, see Eq. 6 in the main manuscript, and can be used to implement a linear PI control loop to stabilize the optical phase as demonstrated in Fig. 3 of the manuscript. Additional reset mechanisms may need to be implemented to cope with the limited range of the phase shifter in case of fiber-based setups, where the phase drift can exceed the phase-shifter range.

Note that for arbitrary waveforms, the condition of constant average power in the overlap regions may be violated, i.e., $U_0(t) \neq$ const., prohibiting the generation of clean error signals. Likewise, there exist waveforms that do not contain spectral components inside the overlap region, resulting in no error signal being generated at all. To overcome these difficulties and generate valid feedback signals for arbitrary waveforms, we may rely on well-defined pilot tones inside the spectral overlap regions instead. By designing these pilot tones to destructively interfere at the main output of the MMI, additional distortions of this approach can be avoided.

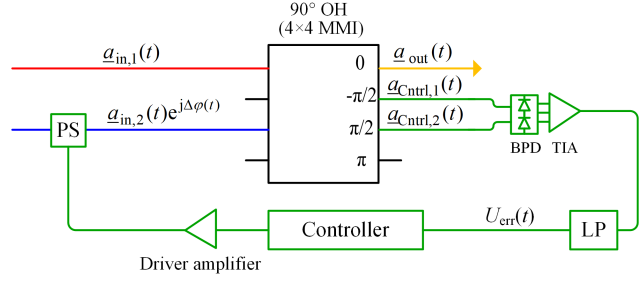

**Fig. S2.** Concept for actively-phase stabilized combining of two overlapping spectral slices using a signal-combining element (SCE) based on a 90° optical hybrid (OH), which may be implemented via a 4×4 multi-mode interference (MMI) coupler. The 2nd and 4th input ports as well as the 4th output port remain unused. The MMI superimposes the input signals with relative phases of $0$, $-\pi/2$, and $\pi/2$ at the various output ports. The two overlapping spectral slices $\underline{a}_{\mathrm{in},1}(t)$ and $\underline{a}_{\mathrm{in},2}(t)\exp(\mathrm{j}\Delta\varphi(t))$ with phase error $\Delta\varphi(t)$ are combined at the "zero-phase" port ("0"), leading to the output signal $\underline{a}_{\mathrm{out}}(t)$. The error signal $U_{\mathrm{err}}(t)$ is generated by feeding the output signals $\underline{a}_{\mathrm{cntrl},1}(t)$ ("$-\pi/2$") and $\underline{a}_{\mathrm{ctnrl},2}(t)$ ("$\pi/2$") to a low-speed balanced photodetector (BPD) followed by a transimpedance amplifier (TIA). Upon low-pass (LP) filtering, the resulting error signal $U_{\mathrm{err}}(t)$ is essentially proportional to the phase error $\Delta\varphi(t)$ for a linear approximation close to the desired operating point $\Delta\varphi = 0$. A proportional-integral (PI) controller is used to drive a phase shifter (PS) that compensates for the measured phase error $\Delta\varphi(t)$.

### 1.2. Error signal generation using 90° optical hybrid

The generation of the error signal $U_{\mathrm{err}}(t)$ for the active phase control based on a 90° optical hybrid as a passive combiner, e.g., implemented as 4×4 MMI, is almost identical to the generation based on a 120° optical hybrid as discussed in the previous section. The setup of the SCE is depicted in Fig. S2.

In analogy to Eq. S2, the three output signals $\underline{a}_{\mathrm{out}}(t)$, $\underline{a}_{\mathrm{cntrl},1}(t)$, and $\underline{a}_{\mathrm{cntrl},2}(t)$ at the output ports 1, 2, and 3 of an ideal 4×4 MMI can be calculated from the input signals $\underline{a}_{\mathrm{in},1}(t)$ and $\underline{a}_{\mathrm{in},2}(t)\exp\{\mathrm{j}\Delta\varphi(t)\}$ connected to the input ports 1 and 3, respectively [1],

$$\begin{bmatrix} \underline{a}_{\mathrm{out}}(t) \\ \underline{a}_{\mathrm{cntrl},1}(t) \\ \underline{a}_{\mathrm{cntrl},2}(t) \end{bmatrix} = \frac{1}{2} \begin{bmatrix} 1 & 1 \\ 1 & -\mathrm{j} \\ 1 & \mathrm{j} \end{bmatrix} \begin{bmatrix} \underline{a}_{\mathrm{in},1}(t) \\ \underline{a}_{\mathrm{in},2}(t)\mathrm{e}^{\mathrm{j}\Delta\varphi(t)} \end{bmatrix} \qquad (S11)$$

Using Eq. S11, the targeted output signal $\underline{a}_{\mathrm{out}}(t)$ as well as the control signal $U_{\mathrm{err}}(t)$ are again calculated in analogy to Eq. S3 and Eq. S4, yielding

$$U_{\mathrm{err}}(t) \approx H_{\mathrm{LP}}(t) * \left[ -\Im\left\{ \underline{a}_{\mathrm{in},1}(t)\underline{a}_{\mathrm{in},2}^{*}(t)\mathrm{e}^{-\mathrm{j}\Delta\varphi(t)} \right\} \right]. \qquad (S12)$$

We use the same assumptions as in the previous section, i.e., that the low-pass filter bandwidth $B_{\mathrm{LP}}$ is, on the one hand, sufficiently large to not distort the phase error $\Delta\varphi(t)$ and, on the other hand, still much smaller than the bandwidth $B_{\mathrm{OR}}$ of the overlap region, $B_{\mathrm{LP}} \ll B_{\mathrm{OR}}$ such that noise and fast modulations of the input signals are suppressed. For stationary signals, we find in accordance to Eq. S10 that $U_{\mathrm{err}}(t) \propto \Delta\varphi(t)$ close to the desired operating point $\Delta\varphi = 0$.

## 2. Calibration

### 2.1. Electro-optic calibration of OAWG transmitter

To compensate linear distortions introduced by the various system components, we pre-distort the drive signals $I_\nu(t)$ and $Q_\nu(t)$ of the various in-phase and quadrature modulators (IQMs), which requires precise knowledge of the associated frequency-dependent transfer functions $\underline{\tilde{H}}_\nu^{(I)}(f)$ and $\underline{\tilde{H}}_\nu^{(Q)}(f)$, $\nu = 1,\ldots,N$, see Eq. 5 in the main manuscript. To this end, we characterize each transmitter channel in a dedicated measurement by means of an IQ receiver (IQR) that was calibrated beforehand using an ultra-stable femtosecond laser (Menhir 1550), see [2–4] for details. The experimental setup for the calibration of the $\nu$-th IQM is shown in

4

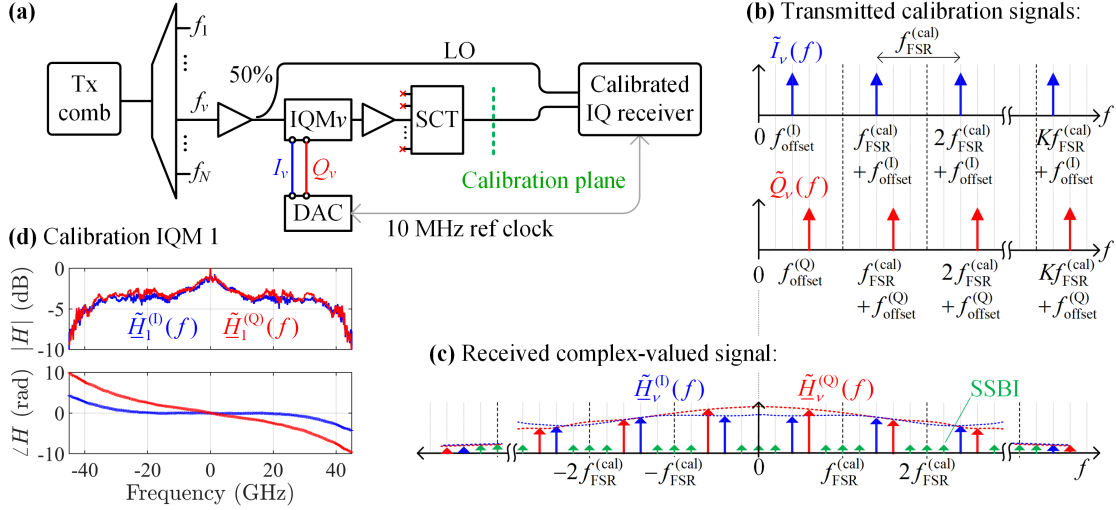

**Fig. S3.** Concept and experimental setup for electro-optic calibration of the OAWG transmitter. **(a)** Experimental setup for calibration of the $\nu$-th transmitter channel. The $\nu$-th comb line at frequency $f_\nu$ of the transmitter comb (Tx comb) is amplified and split equally into a local oscillator (LO) tone, that is fed to a calibrated IQ receiver (IQR), and an optical carrier, that is fed to the IQ modulator (IQM$\nu$) associated with the channel of interest. The in-phase and quadrature inputs of the IQM are driven by the associated radio frequency (RF) calibration signals $I_\nu(t)$ and $Q_\nu(t)$, both of which are designed to exhibit comb-like power spectra, see Subfigure (b). The optical output signal of IQM$\nu$ is amplified, sent through the signal-combining tree (SCT), and received by the calibrated IQR. **(b)** Visualization of the spectra $\tilde{I}_\nu(f)$ (blue) and $\tilde{Q}_\nu(f)$ (red) of the RF calibration signals generated by the digital-to-analog converters (DACs). Note that the two combs are spectrally interleaved to ease the separation of the in-phase and quadrature signal components at the IQR. **(c)** Spectrum of the complex-valued signal received by the calibrated IQR. The transmitter transfer functions $\underline{\tilde{H}}_\nu^{(I)}(f)$ and $\underline{\tilde{H}}_\nu^{(Q)}(f)$ can be extracted by determining the amplitude and phase of the various comb tones and by interpolating for frequencies in between. The spectral position of the calibration comb lines is engineered to avoid distortions by signal-signal beat interference (SSBI, green). **(d)** Amplitude and phase of exemplary transfer functions $\underline{\tilde{H}}_\nu^{(I)}(f)$ and $\underline{\tilde{H}}_\nu^{(Q)}(f)$ for transmitter $\nu = 1$.

Fig. S3 (a). The $\nu$-th comb line at frequency $f_\nu$ is amplified and split into two copies, where one copy is used as an optical carrier and fed to IQM$\nu$ and the other copy is used as local oscillator (LO) tone for homodyne reception. The optical carrier fed to IQM$\nu$ is modulated by the calibration signals $I_\nu(t)$ and $Q_\nu(t)$, and the generated optical waveform is subsequently amplified, propagates though the signal-combining tree (SCT), see Fig. 1 in the main manuscript, and is detected by the calibrated IQR. Note that we use a homodyne detection setup and synchronize the analog-to-digital converters (ADCs) of the calibrated IQR to the digital-to-analog converters (DACs) that drive the IQMs to minimize the impact of optical phase noise at the receiver and to eliminate any frequency offset between the optical carrier and the LO. Because the calibration must include the time delay between the in-phase and quadrature components (IQ-skew) we measure both transfer functions, $\underline{\tilde{H}}_\nu^{(I)}(f)$ and $\underline{\tilde{H}}_\nu^{(Q)}(f)$ simultaneously. To this end, we generate two interleaved radio frequency (RF) frequency combs as calibration signals, where one RF comb is use as a drive signal $I_\nu(t)$ for the in-phase component of IQM$\nu$ whereas the other RF comb serves as a drive signal $Q_\nu(t)$ for the quadrature component. The two combs have the same free spectral range (FSR) $f_{\text{FSR}}^{(\text{cal})}$ but different frequency offsets $f_{\text{offset}}^{(I)}$ and $f_{\text{offset}}^{(Q)}$, see Fig. S3 (b),

$$I_\nu(t) = \sum_{k=0}^{K} \cos\left[2\pi\left(kf_{\text{FSR}}^{(\text{cal})} + f_{\text{offset}}^{(I)}\right)t + \varphi_k^{(I)}\right], \tag{S13}$$

$$Q_\nu(t) = \sum_{k=0}^{K} \cos\left[2\pi\left(kf_{\text{FSR}}^{(\text{cal})} + f_{\text{offset}}^{(Q)}\right)t + \varphi_k^{(Q)}\right]. \tag{S14}$$

In this relation, $K$ is the total number of comb lines of the calibration signal and $\varphi_k^{(I)}$ and $\varphi_k^{(Q)}$ are the initial phases of the various comb lines. The distinct frequency offsets $f_{\text{offset}}^{(I)}$ and $f_{\text{offset}}^{(Q)}$ allow to easily separate the

5

components related to the in-phase and quadrature drive signals at the output of the calibrated IQR, Fig. S3 (c). For each comb-tone frequency, we can directly extract the complex-valued transfer functions $\underline{\tilde{H}}_\nu^{(I)}(f)$ and $\underline{\tilde{H}}_\nu^{(Q)}(f)$ from the amplitude and phase of the received comb lines, indicated in red and blue in Fig. S3 (c), and these data points can then be used to estimate $\underline{\tilde{H}}_\nu^{(I)}(f)$ and $\underline{\tilde{H}}_\nu^{(Q)}(f)$ at any other frequency by interpolation. Exemplary amplitude and phase transfer functions as measured for IQM1 are shown in Fig. S3 (d).

Note that the accuracy of the obtained transfer functions $\underline{\tilde{H}}_\nu^{(I)}(f)$ and $\underline{\tilde{H}}_\nu^{(Q)}(f)$ may be impaired by amplified spontaneous emission (ASE) noise, ADC and DAC noise, signal-signal beat interference (SSBI), and by the accuracy of the IQR calibration. In the following, we will briefly discuss the different impairments and possible mitigation measures:

- **ASE noise:** Amplified spontaneous emission (ASE) noise cannot be fully avoided. However, its effect on the calibration result can be reduced by averaging multiple calibration measurements.

- **ADC and DAC noise:** The root-mean-square (RMS) voltage noise added by an ADC or DAC usually takes a fixed percentage of the full-scale voltage, and the signal is typically adjusted to fill the full input or output range of the ADC or DAC, respectively. As a result, the achievable signal-to-noise ratio (SNR) depends on the peak-to-average power ratio (PAPR) of the signal. To reduce the PAPR and consequently increase the SNR of the transmitted and received signals, the initial phases $\varphi_k^{(I)}$ and $\varphi_k^{(Q)}$ are chosen such that the comb spectra $\tilde{I}_\nu(f)$ and $\tilde{Q}_\nu(f)$ have a quadratic phase profile. This is equivalent to chirping the time domain pulse trains $I_\nu(t)$ and $Q_\nu(t)$ such that the gaps between the periodic pulses vanish and the optical power of the calibration signal is approximately constant. Additionally, ADC and DAC noise can be further reduced by averaging multiple calibration measurements.

- **Signal-signal beat interference (SSBI):** The common-mode rejection ratio (CMRR) of BPDs is generally limited, which results in SSBI distorting the received waveforms and leading to deterministic errors in the retrieved transfer functions $\underline{\tilde{H}}_\nu^{(I)}(f)$ and $\underline{\tilde{H}}_\nu^{(Q)}(f)$. These impairments are systematic and cannot be removed by averaging more measurements. However, a proper choice of the frequency offsets $f_{\text{offset}}^{(I)}$ and $f_{\text{offset}}^{(Q)}$ Fig. S3 (b), allows eliminate the effect of SSBI entirely. For example, by choosing $f_{\text{offset}}^{(I)} = 0.4 \times f_{\text{FSR}}^{(\text{cal})}$ and $f_{\text{offset}}^{(Q)} = 0.6 \times f_{\text{FSR}}^{(\text{cal})}$ as in Fig. S3 (b), the SSBI spectrum at the receiver will only contain beat notes at frequencies $f_{\text{SSBI}} = k f_{\text{FSR}}^{(\text{cal})}$, and $f_{\text{SSBI}} = (k \pm 0.2) f_{\text{FSR}}^{(\text{cal})}$ with integer $k$ (green comb lines in Fig. S3 (c)) and thus not overlap with the desired signal-LO beat notes at frequencies $(k + 0.4) f_{\text{FSR}}^{(\text{cal})}$ (blue comb lines in Fig. S3 (c)) and $(k + 0.6) f_{\text{FSR}}^{(\text{cal})}$ (red comb lines in Fig. S3 (c)), respectively.

- **Calibration errors of the IQR:** The IQM calibration will directly inherit errors from the IQR calibration. Here, we investigate the impact of our receiver calibration on the retrieved transmitter transfer functions $\underline{\tilde{H}}_\nu^{(I)}(f)$ and $\underline{\tilde{H}}_\nu^{(Q)}(f)$ by repeating the transmitter calibration using a different IQR (IQR2). We find that the obtained transfer functions $\underline{\tilde{H}}_{\nu,\text{IQR1}}^{(I)}(f)$, $\underline{\tilde{H}}_{\nu,\text{IQR2}}^{(I)}(f)$, and $\underline{\tilde{H}}_{\nu,\text{IQR1}}^{(Q)}(f)$, $\underline{\tilde{H}}_{\nu,\text{IQR2}}^{(Q)}(f)$ differ only slightly. We quantify the difference by the average relative error $J$,

$$J = \frac{\int_{-B}^{B} \left| \underline{\tilde{H}}_{\nu,\text{IQR1}}^{(I)}(f) - \underline{\tilde{H}}_{\nu,\text{IQR2}}^{(I)}(f) \right|^2 \mathrm{d}f}{\int_{-B}^{B} \left| \underline{\tilde{H}}_{\nu,\text{IQR1}}^{(I)}(f) + \underline{\tilde{H}}_{\nu,\text{IQR2}}^{(I)}(f) \right|^2 \mathrm{d}f}, \tag{S15}$$

which for our transmitter calibrations is of the order of $-35\,\text{dB}$ to $-45\,\text{dB}$ for a bandwidth $B = 45\,\text{GHz}$. Note, however, that both IQRs were calibrated using the same optical reference waveform (ORW), such that systematic errors originating from the ORW or receiver calibration procedure cannot be detected.

- **System nonlinearities:** System nonlinearities cannot be compensated using this approach, because the underlying system model is purely linear, see Eq. 5 in the main manuscript. Nevertheless, in some situations, e.g., when using semiconductor optical amplifiers (SOAs) with non-negligible pattern effects rather than more ideal erbium-doped fiber amplifiers (EDFAs), the calibration accuracy can be improved by choosing IQ calibration signals $I_\nu(t)$ and $Q_\nu(t)$ that have similar statistical properties as

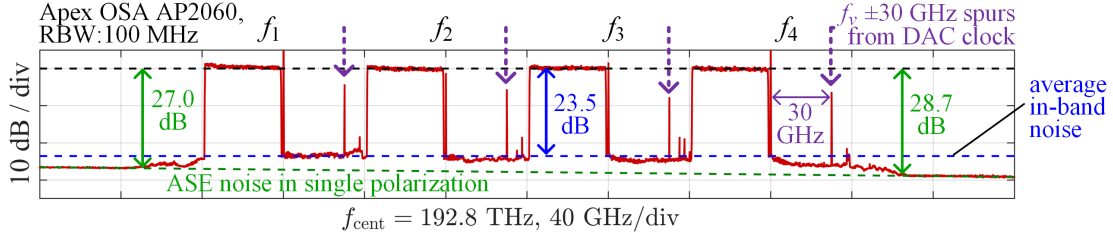

**Fig. S4.** Optical spectrum of a signal generated by the OAWG system and measured after passing through a linear polarizer. The measurement was taken using a high-resolution optical spectrum analyzer (Apex AP2060, resolution bandwidth: 100 MHz). The spectrum consists of four identical single-sideband 37.5 GBd 16QAM signals occupying the low-frequency sideband of each transmitter channel. The configuration allows determining the single-polarization amplified spontaneous emission (ASE) noise (green dashed line) as well as the noise and distortions in the empty upper sidebands (blue dashed line), including ASE noise, spurs and noise from the digital-to-analog converters (DACs), noise from the and electrical amplifiers driving the in-phase and quadrature modulators (IQMs), and IQ crosstalk from the lower sidebands. The spurs originating from the DACs (Keysight M8194A) occur $\pm 30$ GHz away from each optical carrier $f_1, ..., f_4$ and are marked with dashed purple arrows. Note that the spurious tones exist for both sidebands but can only be seen in the empty upper sideband.

the IQ drive used for the target waveform. To this end, the phases $\varphi_k^{(I)}$ and $\varphi_k^{(Q)}$ of the various comb lines can be engineered. For example, choosing the phases $\varphi_k^{(I)}$ and $\varphi_k^{(Q)}$ randomly results in calibration signals $I_\nu(t)$ and $Q_\nu(t)$ with an approximately Gaussian amplitude distribution – similar to the IQ drive signals used for the generation of data signals in our experiments.

### 2.2. Calibration verification and SNR estimation

To validate the transmitter calibration and to measure the in-band noise, we generate single-sideband signals (37.5 GBd 16QAM) using the various IQMs. The resulting optical spectra are recorded by a high-resolution optical spectrum analyzer (OSA, Apex AP2060) at a resolution bandwidth (RBW) of RBW = 100 MHz, see Fig. S4. First, we observe that the spectra are flat and that the positive sideband is well suppressed, indicating a low level of remaining IQ imbalance and thus a good transmitter calibration and bias-point setting of the modulator. Second, we measure the power spectral density of the ASE noise originating from the various optical amplifiers to be on average 27.7 dB below the power spectral density of the single sideband signals, yielding an ASE noise limited signal-to-noise-and-distortion ratio (SNDR) of $\text{SNDR}_{\text{ASE}}^{(4 \times 37.5\,\text{GBd})} = 27.7$ dB. Note that the ASE noise floor is slightly tilted, leading to a higher noise level at lower optical frequencies ($-27.0$ dB) and a lower noise level at higher frequencies ($-28.7$ dB), green dashed line in Fig. S4. As a third step, we estimate the in-band noise and distortions. To this end, we average the residual spectral power in the upper sideband of each channel that should be empty for an ideal single-sideband modulation. A dashed blue line in Fig. S4 indicates the average noise-and-distortion power level. We calculate the associated SNDR, $\text{SNDR}_{\text{Tx}}^{(4 \times 37.5\,\text{GBd})} = 23.5$ dB, see blue double-arrow in Fig. S4. The unwanted spectral components in the upper sidebands include ASE noise, spurs and noise from the DACs, noise from the electrical amplifiers driving the IQMs, as well as residual IQ crosstalk from the lower sidebands. The spurs marked with purple dashed arrows in Fig. S4, occur 30 GHz away from each optical carrier and originate from the DACs (Keysight M8194A).

Based on the in-band noise measurements for the 37.5 GBd 16QAM single sideband signals shown in Fig. S4, we estimate the achievable transmitter SNDR for ultra-broadband QAM data signals generated by the OAWG system according to the experimental setup shown in Fig. 4 in the main manuscript. To this end, we make two assumptions. First, the SCE is treated as an ideal element in the sense that it merges all tributaries with zero phase error. Second, we assume that the PAPR of the IQ drive signals used to generate the 37.5 GBd single-sideband signal (Fig. S4) and the IQ drive signals used to generate the 320 GBd QAM waveform (Fig. 5 in the main manuscript) are similar. The $\text{SNDR}_{\text{Tx}}^{(320\,\text{GBd})}$ of the 320 GBd waveform, which is

7

associated with a 320 GHz noise bandwidth, can hence be estimated from the $\text{SNDR}_{\text{Tx}}^{(4\times 37.5\,\text{GBd})}$, which refers to a noise bandwidth of $4 \times 37.5\,\text{GHz}$ by accounting for the difference in noise bandwidth.

$$\text{SNDR}_{\text{Tx}}^{(320\,\text{GBd})} \approx \text{SNDR}_{\text{Tx}}^{(4\times 37.5\,\text{GBd})} - 10\log_{10}\left(\frac{320\,\text{GHz}}{4\times 37.5\,\text{GHz}}\right)$$
$$\approx 23.5\,\text{dB} - 3.3\,\text{dB} = 20.2\,\text{dB}. \tag{S16}$$

Note that the above estimation may slightly overestimate true transmitter SNDR as we do not record the actual time-domain traces and can therefore not identify all possible signal distortions. Nevertheless, this approach offers the possibility to characterize the signal quality of the OAWG transmitter without including impairments from an IQR as it is, e.g., the case for the results presented in Fig. 5 in the main manuscript.

### 3. Constellation signal-to-noise ratio

The constellation signal-to-noise ratio (CSNR) is used in the main manuscript to evaluate the signal quality. The CSNR is defined as the ratio between the average power $P_S$ of properly normalized ideal symbols with associated complex-valued symbol amplitudes $\underline{S}_{0,n}$ and the average noise-and-distortion power $P_{\text{ND}}$ contained in the actually received symbols with complex-valued symbol amplitude $\underline{S}_n$,

$$\text{CSNR} = \frac{P_S}{P_{\text{ND}}} = \frac{\frac{1}{N_{\text{sym}}}\sum_{n=1}^{N_{\text{sym}}}\left|\underline{S}_{0,n}\right|^2}{\frac{1}{N_{\text{sym}}}\sum_{n=1}^{N_{\text{sym}}}\left|\underline{S}_n - \underline{S}_{0,n}\right|^2}. \tag{S17}$$

In the above relation, $N_{\text{sym}}$ is the number of symbols over which the signal and noise-and-distortion powers are averaged. Often, the CSNR is specified in dB, $\text{CSNR}_{\text{dB}} = 10\log_{10}(\text{CSNR})$. Note that Eq. S17 requires that the received and transmitted complex-valued symbol amplitudes are normalized and rotated correctly. Further, note that the CSNR is the square of the reciprocal the error vector magnitude (EVM) normalized to the average signal power [5–7]. Both quantities can thus be related by

$$\text{EVM}_a = \sqrt{1/\text{CSNR}}. \tag{S18}$$



\end{document}